\documentclass[12pt]{article}
\pdfoutput=1
\usepackage{amsmath,amssymb,graphicx}
\usepackage[T1]{fontenc}
\usepackage[charter]{mathdesign}
\usepackage[font=small,labelfont=bf]{caption}
\usepackage{longtable}
\usepackage{verbatim}
\usepackage{cite}
\usepackage{url}

\usepackage[usenames,dvipsnames]{xcolor}
\definecolor{burgundy}{rgb}{0.565,0.0,0.125}

\usepackage{afterpage}

\newcommand{\beq}{\begin{eqnarray}}
\newcommand{\eeq}{\end{eqnarray}}
\def\OMIT#1{{}}
\newcommand{\lsim}{\mathrel{\rlap{\lower4pt\hbox{\hskip1pt$\sim$}}
    \raise1pt\hbox{$<$}}}         
\newcommand{\gsim}{\mathrel{\rlap{\lower4pt\hbox{\hskip1pt$\sim$}}
    \raise1pt\hbox{$>$}}}         

\usepackage{titlesec}

\titleformat{\section}
{\color{gray}\normalfont\Large\bfseries}
{\color{gray}\thesection}{1em}{}

\titleformat{\subsection}
{\color{gray}\normalfont\large\bfseries}
{\color{gray}\thesubsection}{1em}{}

\title{\bf \color{gray} Naturalness, $b \to s\gamma$, and SUSY Heavy Higgses}

\author{Andrey Katz, Matthew Reece, and Aqil Sajjad\\
{\small \color{gray} \texttt{andrey, mreece, sajjad@physics.harvard.edu}}\\
{\em Department of Physics, Harvard University, Cambridge, MA 02138, USA}}

\begin{document}
\maketitle

\begin{abstract}
\noindent 
We explore naturalness constraints on the masses of the heavy Higgs bosons $H^0, H^\pm$, 
and $A^0$ in supersymmetric theories.
We show that, in any extension of MSSM which accommodates the 125~GeV Higgs 
at the tree level, one can derive an upper bound on the SUSY 
Higgs masses from naturalness considerations. As is well-known for the MSSM, these bounds become weak at large $\tan \beta$. However, we show that measurements of $b \to s \gamma$ together with naturalness arguments lead to an upper bound on $\tan \beta$, strengthening the naturalness case for heavy Higgs states near the TeV scale. The precise bound depends somewhat on the SUSY mediation scale: allowing a factor of 10 tuning in the stop sector, the measured rate of $b \to s\gamma$ implies $\tan \beta \lsim 30$ for running down from 10 TeV but $\tan \beta \lsim 4$ for mediation at or above 100 TeV, placing $m_A$ near the TeV scale for natural EWSB. Because the signatures of heavy Higgs bosons at colliders are less susceptible to being ``hidden'' than standard superpartner signatures, there is a strong motivation to make heavy Higgs searches a key part of the LHC's search for naturalness. In an appendix we comment on how the Goldstone boson equivalence theorem links the rates for $H \to hh$ and $H \to ZZ$ signatures.
\end{abstract}

\section{Introduction}

The most compelling argument for the possibility of supersymmetry near the weak scale is that it allows for the 
possibility of {\em natural} electroweak symmetry breaking. This possibility, however, hinges on a number of 
conditions~\cite{Barbieri:1987fn,Dimopoulos:1995mi,Pomarol:1995xc,Cohen:1996vb}. 
The tree-level electroweak symmetry breaking conditions show that the Higgs VEV should not be much 
larger than the higgsino mass parameter $\mu$, so naturalness requires light 
higgsinos~\cite{Barbieri:1987fn,Kitano:2006gv,Perelstein:2007nx,Papucci:2011wy,Brust:2011tb,Baer:2013gva}. 
At one loop, the top quark correction to the up-type Higgs mass parameter $m_{H_u}^2$ must be approximately 
canceled, requiring light stop 
squarks~\cite{Meade:2006dw,Kitano:2006gv,Perelstein:2007nx,Papucci:2011wy,Brust:2011tb,Farina:2013ssa,Kribs:2013lua,Kowalska:2013ica,Han:2013kga}. Finally, the stops themselves suffer from large corrections due to a gluino loop, requiring that the gluinos must also not be too heavy~\cite{Papucci:2011wy,Brust:2011tb}. Although higgsinos are difficult to constrain experimentally, the search for natural SUSY has driven an extensive effort to discover stops or gluinos at the LHC. This effort has succeeded in placing stringent bounds on their possible masses and decay modes. For an up-to-date review of the implications of LHC data for SUSY, see Ref.~\cite{Craig:2013cxa}. (Also see refs.~\cite{Feng:2013pwa} and~\cite{Hardy:2013ywa} for recent overviews of the status of SUSY naturalness.)

When we discuss natural SUSY, we will always have in mind a scenario with relatively light stops. 
Loop corrections from the stops in natural scenarios are not sufficient to lift the SM-like Higgs mass to 125 GeV. As a result, we will assume that new physics {\em beyond} the MSSM provides a new contribution to the Higgs quartic coupling 
and raises the Higgs mass at the tree level. Many options are available for this, including new $F$ or $D$-term quartics~\cite{Batra:2003nj,Barbieri:2006bg,Dine:2007xi}.

In this paper we explore to what extent the heavy Higgs bosons of the MSSM and its extensions, $H^0$, $H^\pm$, and $A^0$, could also constitute probes of naturalness. As with the higgsino mass parameter $\mu$, their mass terms appear in the tree-level conditions for EWSB, so naturalness will not be consistent with arbitrary values of these parameters. The reason that heavy Higgses have not joined the usual pantheon of naturalness signatures is that, in the MSSM, it is only the ratio $m_A/\tan \beta$ of their mass scale to $\tan \beta$ that is constrained, so at large $\tan \beta$ they can be out of reach of colliders without requiring any fine-tuning~\cite{Perelstein:2007nx}. On the other hand, in scenarios like $\lambda$SUSY where $\tan \beta$ is order-one~\cite{Barbieri:2006bg}, it is known that naturalness requires the other Higgs bosons to be light~\cite{Hall:2011aa}. The tuning cost of raising the heavy Higgs masses when $\tan\beta$ is not large was recently emphasized in Ref.~\cite{Gherghetta:2014xea}.

Our goal in this paper is to construct an argument that, in any given extension of the MSSM, even at large $\tan\beta$, there is an upper bound on the heavy Higgs masses arising from naturalness. We derive simple expressions for the fine-tuning when different possible quartic couplings are added to raise the Higgs mass to 125 GeV. The only case in which there is not an immediate bound is the MSSM-like case of an $\left|H_u\right|^4$ quartic, for which the heavy Higgses can be made heavy while simultaneously going to large $\tan \beta$. However, we will argue that measurement of $b \to s\gamma$, together with a combination of naturalness and direct constraints on other superpartner masses, allows us to cut off the large-$\tan\beta$ tail of the natural parameter space. The fact that $b \to s\gamma$ is difficult to suppress in natural SUSY due to a contribution from a loop of stops and higgsinos was emphasized in ref.~\cite{Ishiwata:2011ab}. Our discussion will be somewhat more general because we assume that the Higgs mass is lifted by quartic couplings beyond the MSSM, relaxing constraints on $A_t$ assumed in that reference. Nonetheless, we will find a constraint.

Thus, there is a bound from a combination of tree-level naturalness and $b \to s\gamma$ measurements on the mass scale of heavy Higgs bosons in a natural supersymmetric theory. Unlike the tree-level constraint on higgsinos, which is difficult to exploit because they can be essentially invisible at colliders, the parameter space for natural heavy Higgses can be significantly constrained by data. Both direct searches and ${\cal O}(v^2/m_H^2)$ corrections to the light Higgs boson decay widths play a part in this. We close our paper with a brief look at the prospects for experimental tests of natural SUSY in these heavy Higgs search channels.

\section{Tree-level fine-tuning}
\label{sec:treelevelFT}

The most general renormalizable potential for the two Higgs doublets $H_u$ and $H_d$ is~\cite{Gunion:2002zf,Randall:2007as,Blum:2012kn}:
\beq
V(H_u, H_d) & = &  M_u^2 \left|H_u\right|^2 + M_d^2 \left|H_d\right|^2 + \left(b H_u \cdot H_d + {\rm h.c.}\right) + \frac{1}{4} \lambda_1 \left|H_u\right|^4 + \lambda_2 H_u^\dagger H_u \left(H_u \cdot H_d + {\rm h.c.}\right) \nonumber \\
& & + \lambda_3 \left|H_u\right|^2 \left|H_d\right|^2 + \frac{1}{2} \lambda_4 \left(H_u \cdot H_d + {\rm h.c.}\right)^2 + \lambda_5 \left|H_u \cdot H_d\right|^2 \nonumber \\
& & + \lambda_6 H_d^\dagger H_d  \left(H_u \cdot H_d + {\rm h.c.}\right)  + \frac{1}{4} \lambda_7 \left|H_d\right|^4.
\label{eq:higgspotential}
\eeq
Here $H_u \cdot H_d$ denotes the SU(2)-invariant contraction with an antisymmetric $\varepsilon$ symbol. One may be tempted to write another term $(H_d^\dagger H_u)(H_u^\dagger H_d)$, but this is just the linear combination $\left|H_u\right|^2 \left|H_d\right|^2 - \left|H_u \cdot H_d\right|^2$ and can be absorbed into $\lambda_3$ and $\lambda_5$. For simplicity we use the notation $M_u^2 \equiv \left|\mu\right|^2 + m_{H_u}^2$ and $M_d^2 \equiv \left|\mu\right|^2 + m_{H_d}^2$. In the MSSM, the nonzero tree-level quartic couplings are: 
\beq
\lambda_1 = \lambda_7 = \frac{g^2 + g'^2}{2};~~\lambda_3 = \frac{g^2 - g'^2}{4};~~\lambda_5 = -\frac{g^2}{2}.
\label{eq:Dtermquartics}
\eeq
However, in the MSSM at tree level the Higgs mass is always smaller than the measured value, so we must raise it. For the most part, in this paper, we will simply assume that the Higgs mass is lifted by a new, hard SUSY-breaking contribution to one of the quartic couplings $\lambda_i$, and that beyond-MSSM physics otherwise does not affect the Higgs potential. The new term could arise from new $F$-terms in higher-dimension operators~\cite{Dine:2007xi,Lu:2013cta} or from nondecoupling $D$-terms from new gauge groups~\cite{Batra:2003nj,Maloney:2004rc,Cheung:2012zq}. 

In some cases, the detailed physics lifting the Higgs mass will also affect Higgs properties in more significant ways, e.g. when mixing with a singlet~\cite{Hall:2011aa,Espinosa:1991gr,Agashe:2012zq,Gherghetta:2012gb} or triplet~\cite{Dine:2007xi,Agashe:2011ia} is important. We will not consider these models in detail, but we expect that although they may provide further experimental search channels they will not alter the basic conclusion about whether decoupling the heavy Higgs bosons is natural.

In this section we will focus on the quartic couplings $\lambda_1 \left|H_u\right|^4$ and $\lambda_5 \left|H_u \cdot H_d\right|^2$, which we view as well-motivated possibilities. We will not discuss the other cases, but a similar exercise can be carried out for all of them. The quartics $\lambda_6$ and $\lambda_7$ have effects only at small $\tan \beta$, which is disfavored because it requires a very large top Yukawa coupling. The couplings $\lambda_3$ and $\lambda_4$ have a similar effect to $\lambda_5$, since they involve two up-type Higgs bosons and two down-type Higgs bosons. The coupling $\lambda_2$ is an interesting intermediate case, favoring moderate $\tan \beta$, but we don't know of a model in which in dominates.

\subsection{Reminder: EWSB and tuning in the tree-level MSSM}

Given the potential in eq.~\eqref{eq:higgspotential}, we can vary with respect to the VEVs of $H_u^0$ and $H_d^0$ to obtain the conditions for an electroweak symmetry breaking vacuum of VEV $v$. These equations, for the case of the MSSM, are:
\beq\label{eq:mssmeom1}
M_U^2 & = & b \cot \beta + \frac{1}{2} m_Z^2 \cos(2\beta) \\ \label{eq:mssmeom2}
M_D^2 & = & b \tan \beta - \frac{1}{2} m_Z^2 \cos(2\beta).
\eeq
The appearance of $m_Z^2$ here comes from assuming that only the tree-level $D$-term quartic couplings are present. Of course, this assumption is not consistent with the observed Higgs mass in our universe, since $m_h^2 < m_Z^2$ in the tree-level MSSM. Nonetheless, it is useful to take a quick look at tuning in this case because it is familiar and it offers a useful starting point before proceeding to theories with more general quartic terms. Adding the two EWSB equations gives
\beq
M_U^2 + M_D^2 = \frac{2 b}{\sin(2\beta)} = m_A^2,
\eeq
using the result one obtains by diagonalizing the pseudoscalar mass matrix. 
On the other hand, multiplying  Eq.~\eqref{eq:mssmeom1}
 by $\tan^2\beta$ and subtracting from Eq.~\eqref{eq:mssmeom2}, we obtain:
\beq
\frac{1}{2} m_Z^2 = \frac{M_D^2 - M_U^2 \tan^2 \beta}{\tan^2 \beta - 1}.
\eeq
In order to have a theory that is not fine-tuned, we would like the individual terms on the right-hand side to be not much larger than the terms on the left-hand side. Recalling that $M_U^2 = m_{H_u}^2 + \left|\mu\right|^2$, we can extract three conditions:
\beq
\left|\mu\right|^2 & \lsim & m_Z^2 \nonumber \\
\left|m_{H_u}^2\right| & \lsim & m_Z^2 \nonumber \\
m_{H_d}^2 & \lsim & m_Z^2 \tan^2 \beta.
\eeq
The first of these equations is the very familiar condition that higgsinos should not be much heavier than the $Z$ boson to prevent tree-level tuning~\cite{Barbieri:1987fn,Perelstein:2007nx,Papucci:2011wy,Brust:2011tb,Baer:2013gva,Baer:2014ica}. The second is unsurprising, 
since $\tan \beta > 1$ so that the Higgs that gets a VEV has a significant component in $H_u^0$. 
In order to obtain a VEV at the weak scale, this Higgs should have a mass near the weak scale. The final condition receives the least attention, although it has been discussed at times in the literature (e.g.~\cite{Perelstein:2007nx,Gherghetta:2014xea}). It tells us that the down-type Higgs soft mass---which, at large $\tan \beta$, is approximately a measure for the mass of the states $A^0, H^0$, and $H^\pm$---cannot be much larger than $m_Z \tan \beta$. The reason this bound typically receives less attention is that it is usually assumed that 
$\tan \beta$ can naturally be very large, allowing the heavy Higgs bosons to be very heavy without a 
large amount of fine-tuning. We think that it is timely to revisit this tree-level naturalness constraint for two reasons. First, many of the models that are frequently studied as ways of lifting the Higgs mass to 125 GeV operate best at small-to-moderate $\tan\beta$. Second, we will argue that the measurement of $b \to s \gamma$ prevents theories with very large values of $\tan \beta$ from being natural. Given such an upper bound on $\tan \beta$, a fine-tuning argument can then impose an upper bound on $m_A$ as well. One of the main goals of this paper is to quantify this upper bound: given what we now know about $b \to s\gamma$, how heavy can the other Higgs bosons be without fine-tuning?

The fine-tuning of EWSB is typically measured in terms of the variation of either the Higgs VEV~\cite{Barbieri:1987fn,Perelstein:2007nx} or the soft mass $m_{H_u}^2$~\cite{Kitano:2006gv,Papucci:2011wy} with respect to the input parameters. We will mostly follow the first, Barbieri-Giudice, 
definition to quantify the tuning of the Higgs VEV with respect to a parameter $x$:
\beq
\Delta_x \equiv \left| \frac{\partial \log v^2}{\partial \log x} \right|.
\eeq
If $\Delta_x \gg 1$ for some parameter $x$, we will say that the theory is fine-tuned. This actually measures the {\em sensitivity} of the Higgs VEV (or, equivalently, the $Z$ mass) to an underlying parameter. We generally think of tuning as occurring when there must be a cancelation between different contributions, requiring a delicate adjustment of different input parameters with respect to one another to achieve a result near the experimentally observed value. For recent discussions of definitions of tuning and how this computation may not always reflect what we think of as fine-tuning, see refs.~\cite{Fichet:2012sn,Baer:2013gva,Fan:2014txa}. We expect that the Barbieri-Giudice measure is typically a fairly good, albeit imperfect, proxy for our intuitive notions of tuning.

We will now explore the tuning measure in various extensions of the MSSM. In each case, we will assume that a particular new hard-SUSY-breaking quartic coupling has been added. We will not worry much about the details of the UV completion, which we expect to have only a mild effect on the fine-tuning bounds that we infer.\footnote{In some examples, e.g. quartic which is generated from non-decoupling D-terms, one would often need an additional fine tuning beyond the MSSM to produce the 
desired quartic. In the D-term scenario this is related to the fact that that the non-decoupling D-term is proportional to the 
soft masses of heavy $W'/Z'$-inos. Of course we do not take this potentially model-dependent 
tuning into account in our analysis. However one should 
bear in mind that the fine tuning that we estimate is the lowest possible bound within the low energy effective theory.}

\subsection{The $\lambda_1 \left(H_u^\dagger H_u\right)^2$ extension}

First we will assume that the new quartic coupling that has been added is dominantly up-type. The usual loop corrections in the MSSM obtained by integrating out stops~\cite{Haber:1990aw,Barbieri:1990ja,Casas:1994us,Carena:1995bx} are of this form. It could also arise from new $D$-terms in conjunction with other 
quartics~\cite{Batra:2003nj,Maloney:2004rc,Craig:2011yk,Craig:2012bs,Cheung:2012zq,Bharucha:2013ela}; its effects would dominate over those of the other quartics at large $\tan \beta$. Finally, we could consider a new source of tree-level $F$-terms by adding a new triplet $T$ with appropriate hypercharge to allow an $H_u \cdot T H_u$ Yukawa coupling, and pairing $T$ with a vectorlike partner $\overline{T}$ with a supersymmetric mass term~\cite{Dine:2007xi,Agashe:2011ia}. For now we will remain agnostic about the UV completion, simply assuming that such a quartic is present in the potential.

\begin{figure}[h]
\begin{center}
\includegraphics[width=0.8\textwidth]{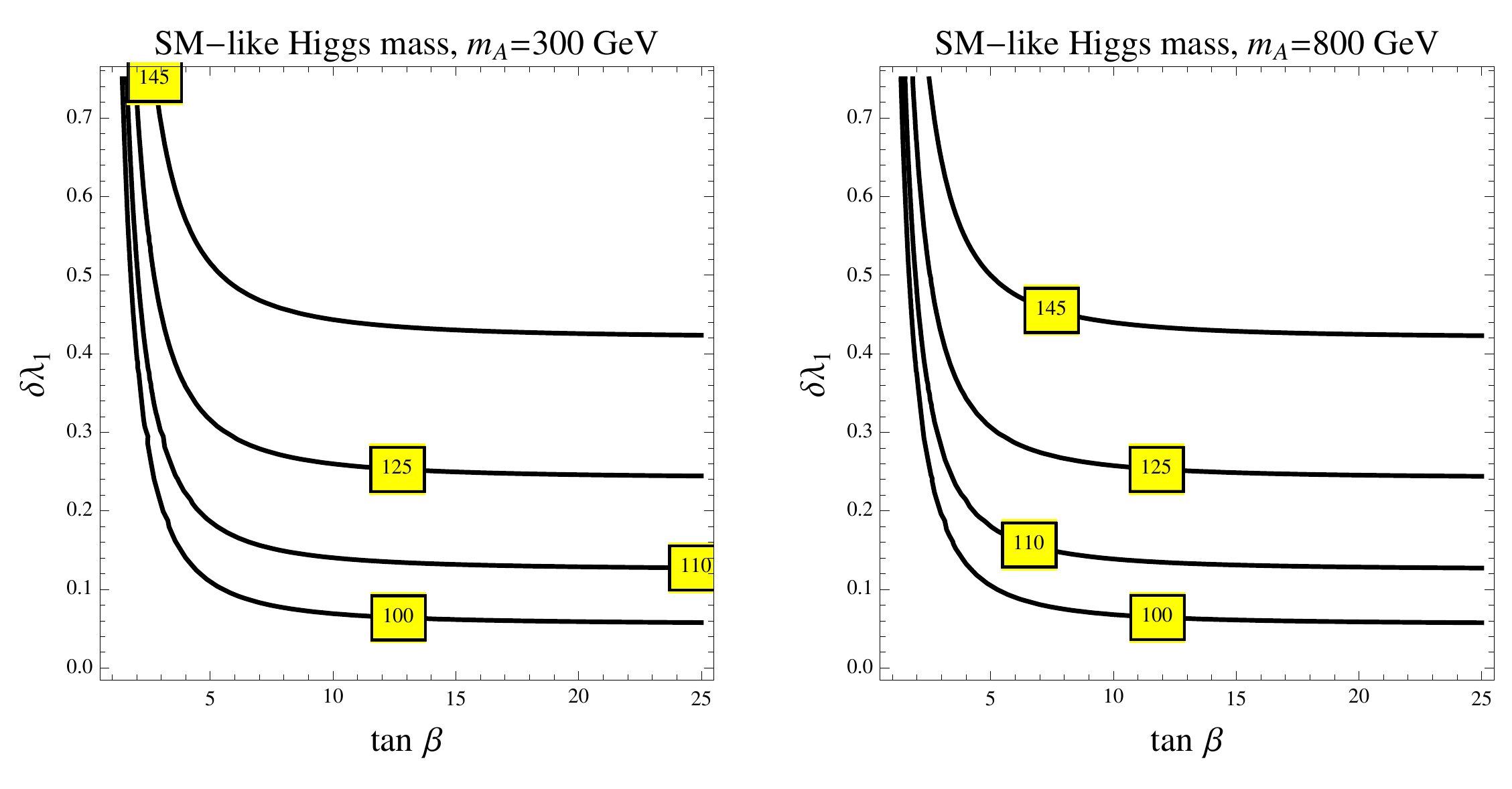}
\end{center}
\caption{Contours of lifted Higgs mass when adding a new $\left|H_u\right|^4$ quartic coupling $\delta \lambda_1$, for two different choices of $m_A$. As expected from eq.~\eqref{eq:dmhdl1}, the dependence on $m_A$ is small. For large $\tan \beta$ we need $\delta \lambda_1 \approx 0.24$ to lift the Higgs mass.
} \label{fig:deltalambda1quartic}
\end{figure}

Given a new contribution $\delta \lambda_1$ to the up-type Higgs quartic (beyond the $D$-term contribution in eq.~\eqref{eq:Dtermquartics}), the two EWSB equations become
\beq
M_U^2 & = & b \cot \beta + \frac{1}{2} m_Z^2 \cos(2\beta) - \frac{\delta \lambda_1}{2} v^2 \sin^2 \beta, \nonumber \\
M_D^2 & = & b \tan \beta - \frac{1}{2} m_Z^2 \cos(2\beta).
\eeq
The new quartic term shifts the mass of the light scalar Higgs eigenstate. The full analytic formula is not very enlightening, but in the $\lambda \ll 1$, $m_A \gg m_Z$ limit we can expand it as:
\beq
\delta m_h^2 & = & \begin{cases}
                                 \delta\lambda_1 v^2 \left(1 - \frac{2}{\tan^2\beta}\left(1 + 2 \frac{m_Z^2}{m_A^2}\right) + \cdots \right), & \text{if $\tan \beta \gg 1$}.\\
                                 \frac{1}{4}\delta\lambda_1 v^2  \left(1 + \left(\tan\beta - 1\right)\left(2 - 2\frac{m_Z^2}{m_A^2}\right) + \cdots \right), & \text{if $\tan \beta \approx 1$}.
                            \end{cases}
\label{eq:dmhdl1}                            
\eeq
Alternatively, for any $\tan \beta$ we can expand the mass formula for $m_A^2 \gg m_Z^2$ as 
\beq
m_h^2 \approx m_Z^2 \cos^2(2\beta) + \delta \lambda_1 v^2 \sin^4 \beta - \frac{\left(2 m_Z^2 \cos 2 \beta - \delta \lambda_1 v^2 \sin^2 \beta\right)^2 \sin^2(2\beta)}{4 m_A^2} + {\cal O}(m_Z^6/m_A^4). \label{eq:mhlambda1}
\eeq
We show the contours of the lifted Higgs mass as a function of $\delta \lambda_1$ and $\tan \beta$ in Fig~\ref{fig:deltalambda1quartic}. 

As we expect for a term involving only the up-type Higgs, the new quartic is more efficient at raising the light Higgs boson mass to 125 GeV in the limit of large $\tan\beta$. Furthermore, because the MSSM tree-level contribution is suppressed at small $\tan \beta$, it becomes even more difficult to obtain $\delta \lambda_1$ large enough in that case. This is illustrated by the contours of constant Higgs mass in the $\left(\tan \beta, \delta \lambda_1\right)$ plane in Figure~\ref{fig:deltalambda1quartic}.

\begin{figure}[h]
\begin{center}
\includegraphics[width=.5\textwidth]{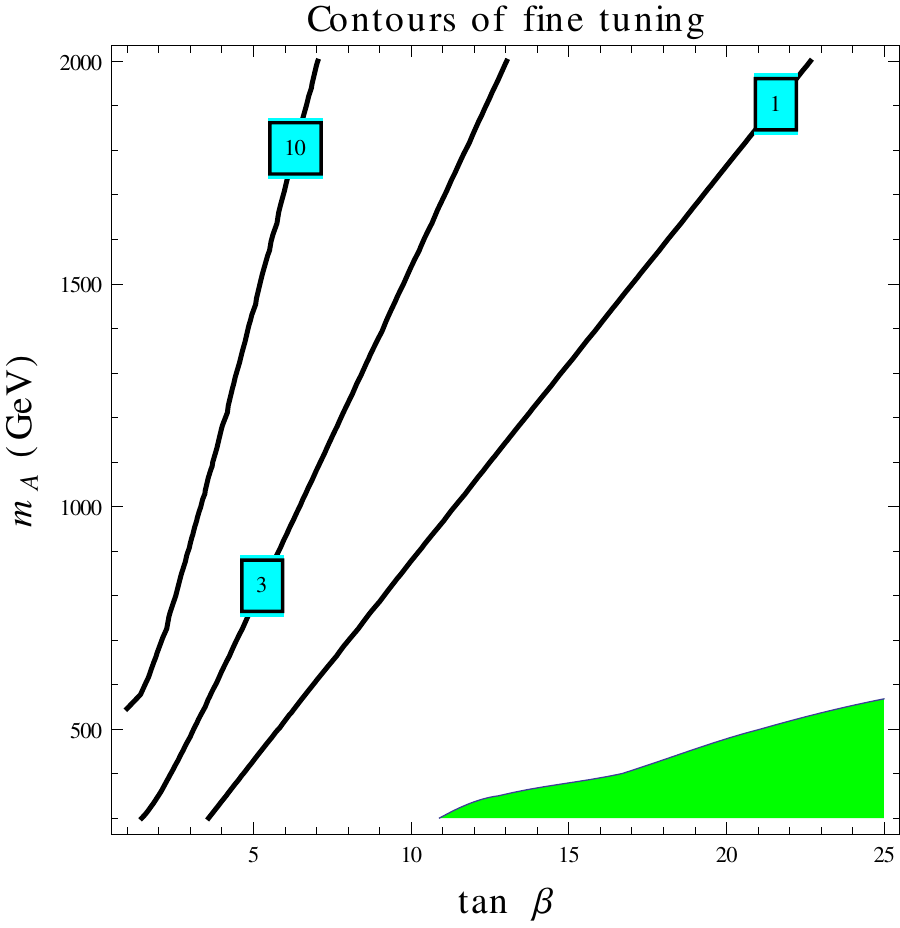}
\end{center}
\caption{Contours of fine tuning of EWSB with an extra quartic $|H_u|^4$. The exact value of $\lambda_1$ is determined by demanding 
$m_h = 125$~GeV. The shaded green region is directly excluded by the CMS search for $H \to \tau^+ \tau^-$ decay (see text
for explanation).  
}
\label{fig:lambda1-FT}
\end{figure}

To evaluate the tuning, we first take a derivative with respect to $M_d^2$. We use the fact that $b = m_A^2 \sin \beta \cos \beta$ (a result that is unchanged from the MSSM case) and thus that $M_d^2$ can be written in terms of the physical parameters $m_Z^2, m_A^2,$ and $\tan \beta$ as $M_d^2 = m_A^2 \sin^2 \beta - \frac{1}{2} m_Z^2 \cos(2\beta)$. The resulting expression is:
\beq
\left|\frac{\partial \log v^2}{\partial \log M_d^2}\right| = \frac{\cos^2 \beta \left[2 m_A^2 + m_Z^2\left(1 - \cot^2\beta\right)\right]\left[m_A^2 \csc^2 \beta + 2 m_Z^2 + \delta \lambda_1 v^2\right]}{m_A^2 \left(m_Z^2 + \delta \lambda_1 v^2\right) + m_Z^2 \cot^2\beta \left[m_A^2 \left(\cot^2 \beta -2\right) + \delta \lambda_1 v^2\right]}.
\eeq
This expression is not very enlightening on its own, but the main question we are interested in is: if we allow at most a given amount of fine-tuning, can we infer a bound on the physical masses of heavy particles? For this question, it is reasonable to expand the tuning measure at large $m_A^2$. We will also assume that the value of $\delta \lambda_1$ is chosen to fix the Higgs mass $m_h^2$ as in eq.~\eqref{eq:mhlambda1}. The result is:
\beq
\left|\frac{\partial\log v^2}{\partial \log M_d^2}\right| & \approx & \frac{1}{2} \sin^2 2\beta \frac{m_A^2 + m_h^2}{m_h^2} + \frac{1}{2} \cos^2\beta \left(1 - 4 \cos 2\beta + \cos 4\beta\right) \frac{m_Z^2}{m_h^2} + {\cal O}(m_h^2/m_A^2) \nonumber \\
& \longrightarrow_{\tan \beta \to \infty} & \frac{2 m_A^2 + 2 m_h^2 + 3 m_Z^2}{m_h^2 \tan^2 \beta}.
\eeq
From this we can see that if $m_A^2 \gg m_h^2$, the theory becomes very fine-tuned unless $\sin(2\beta)$ is small, which happens in the $\tan \beta \gg 1$ limit. We explicitly illustrate this point in Fig.~\ref{fig:lambda1-FT} where we plot the contours of fine tuning as a function of $m_A$ and $\tan \beta$. The precise 
value of $\delta \lambda_1$ on this plot is set by demanding $m_h = 125$~GeV. 

As expected, one gets very low fine tuning for very large values of $\tan \beta$. Even now the large $\tan \beta $ region can be partially explored by the LHC, due to a robust 
$H^0,~A^0 \to \tau^+ \tau^-$ decay mode which  can be directly probed. In Fig.~\ref{fig:lambda1-FT} we show a green region, which 
has been directly excluded by the CMS search~\cite{CMS-PAS-HIG-13-021} for $H^0 \to \tau^+ \tau^-$. 
We anticipate that much more significant gains will be 
made by LHC14.

However, another important constraint on the large $\tan \beta$ region comes from the measurement of the flavor-violating decay $b \to s \gamma$, which we will explore in detail in Section~\ref{sec:flavor}. There we will find that, for {\em very} low-scale SUSY breaking (mediated at $\Lambda = 10$ TeV), one can accommodate $\tan \beta \approx 30$ if one allows a factor of 10 tuning in the stop sector. (Indirect constraints from Higgs decays already force us to accept a minimum factor of about 5 tuning in the stop sector~\cite{Fan:2014txa}.) Using the formula above, we find that if we allow at most an {\em additional} factor of 10 tuning in EWSB, for a combined 1\% tuning, we have $m_A \lsim 8.4$ TeV. Probing such large values of $m_A$ will require future hadron colliders, more powerful than the LHC. On the other hand, we will find in Section~\ref{sec:bsgammaboundresults} that with even a slightly higher mediation scale $\Lambda = 30$ TeV the bound from $b \to s\gamma$ becomes notably stronger: $\tan \beta \lsim 10$. In this case, allowing for at most an additional factor of 10 tuning in EWSB implies $m_A \lsim 2.8$ TeV. If we view the factor of 10 tuning in the stop sector as already deviating from naturalness, and want to ask for no {\em additional} tuning in EWSB, we have the stronger condition $m_A \lsim 0.9$ TeV. Furthermore, higher mediation scales only strengthen the $\tan \beta$ upper bound from $b\to s\gamma$, so although parts of the natural parameter space may require future colliders to probe, in much of the parameter space the heavy Higgs bosons should be accessible at the LHC. Measurements of the light Higgs boson decay modes at the 14 TeV LHC with 300 fb$^{-1}$ of data will probe the range of $m_A$ up to about 450 GeV~\cite{Gupta:2012fy}. Heavier masses can be probed only by direct searches or higher precision measurements at the high luminosity LHC or especially future $e^+ e^-$ colliders.

\subsection{The $\lambda_5 \left|H_u \cdot H_d\right|^2$ extension}

This is the quartic extension that arises in the NMSSM or $\lambda$SUSY. It does not change the pseudoscalar mass relation $m_A^2 = 2b/\sin(2\beta)$. In this case, we find that the light Higgs mass is corrected as:
\beq
m_h^2 = m_Z^2 \cos^2(2\beta) + \delta \lambda_5 v^2 \sin^2(2\beta) - \frac{\left(m_Z^2 - \delta \lambda_5 v^2\right)^2\sin^2(4\beta)}{4m_A^2} + {\cal O}(m_Z^6/m_A^4). \label{eq:mhlambda5}
\eeq
In this case moderate values of $\tan \beta$ are most effective for raising the Higgs mass, because the correction term involves $v_d$ and is 
suppressed in the large $\tan\beta$ limit. In fact, it is impossible to get $m_h = 125$~GeV with large $\tan \beta$. On top of that, we often need
large, almost non-perturbative values of $\delta \lambda_5$ in order to get the correct value of the SM-like Higgs mass.

\begin{figure}[h]
\begin{center}
\includegraphics[width=0.8\textwidth]{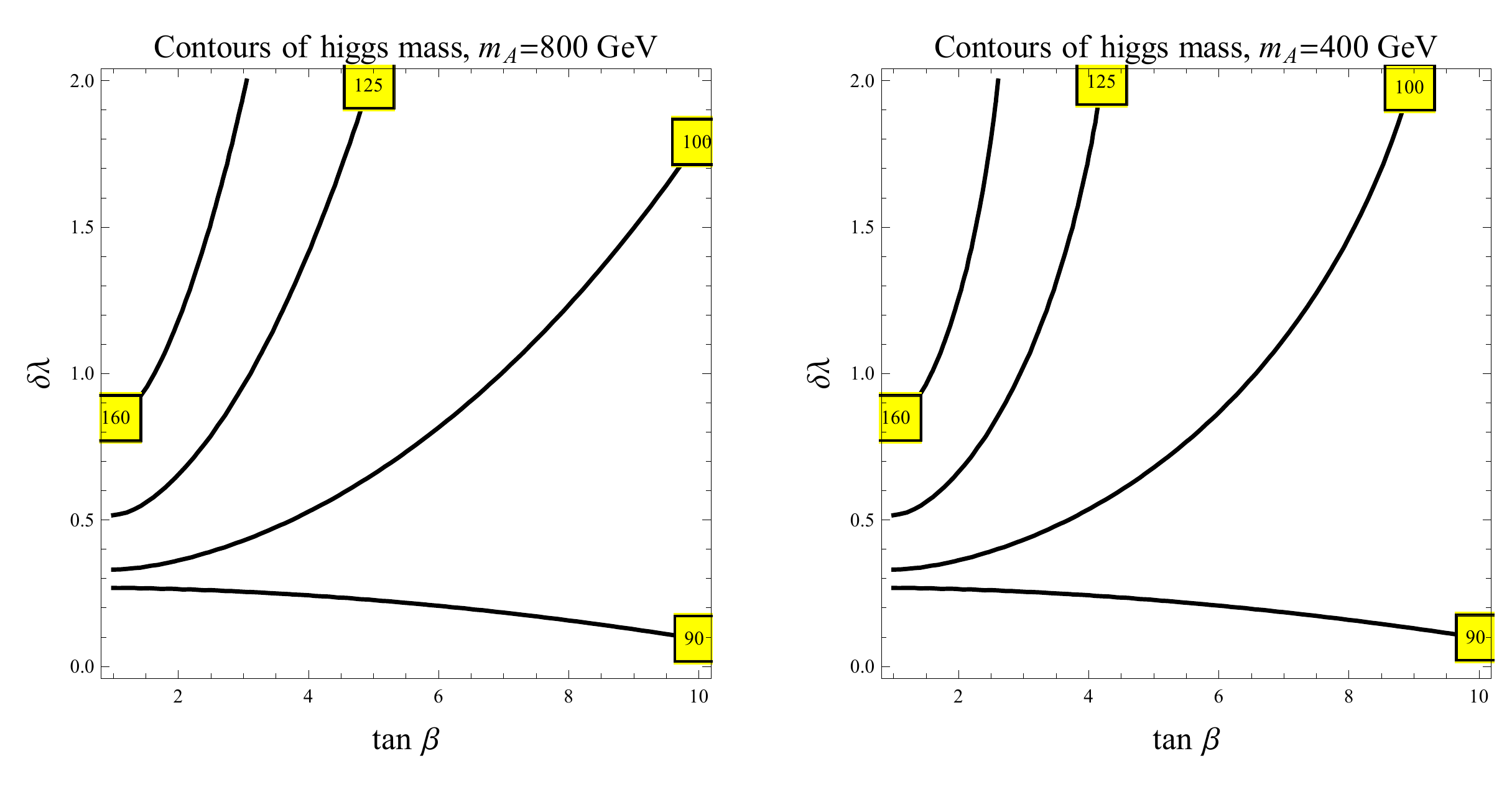}
\end{center}
\caption{Contours of lifted Higgs mass when adding a new $\left|H_u\cdot H_d\right|^2$ quartic coupling $\delta \lambda_5$, 
for two different choices of $m_A$. Only moderate $\tan \beta$ values are allowed by 125~GeV Higgs.
}
\label{fig:deltalambda5quartic}
\end{figure}

The tuning measure in this case is:
\beq
\left|\frac{\partial \log v^2}{\partial \log M_d^2}\right| = \frac{M_d^2 \sin^2(2\beta) \left(m_A^2 \csc^2\beta + 2 m_Z^2 - 2 \delta \lambda_5 v^2\right)}{m_A^2 m_Z^2 + \left(m_A^2 - \delta \lambda_5 v^2\right)\left(m_Z^2 - \delta \lambda_5 v^2\right) \cos(4\beta) + \delta \lambda_5 \left(m_A^2 + m_Z^2\right) v^2 - \delta \lambda_5^2 v^4}.
\eeq
Again, this expression simplifies in the limit $m_A^2 \gg m_h^2, m_Z^2$, choosing $\delta \lambda_5$ to fix the Higgs mass $m_h^2$ as in eq.~\eqref{eq:mhlambda5}:
\beq \nonumber 
\left|\frac{\partial\log v^2}{\partial \log M_d^2}\right| & \approx  & \frac{1}{2} \sin^2 2\beta \frac{m_A^2}{m_h^2} - \frac{3 - 2\cos(2\beta) + 
\cos(4\beta)}{4} + \\
&& \frac{m_Z^2  \left(1 - 4 \cos(2\beta) + \cos(4\beta)\right)}{4 m_h^2}+ {\cal O}(m_{Z,h}^2/m_A^2).
\eeq

\begin{figure}[h]
\begin{center}
\includegraphics[width=.5\textwidth]{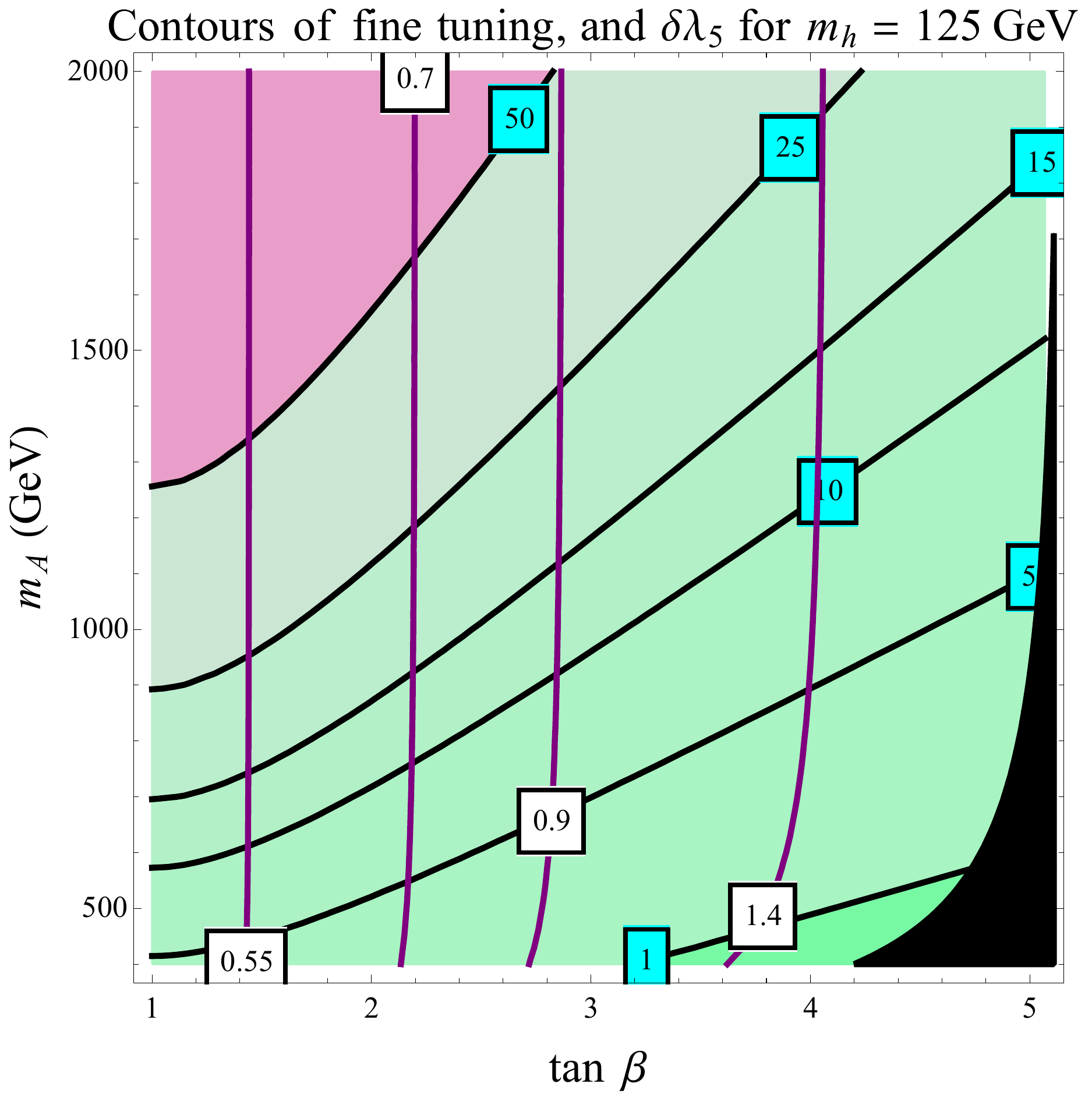}
\end{center}
\caption{Contours of fine tuning of EWSB when adding a new $\left|H_u\cdot H_d\right|^2$ coupling $\delta \lambda_5$. Most of the parameter
space is already fine tuned. The purple contours denote the $\delta \lambda_5$ value needed to get a 125~GeV SM-like Higgs mass. In the black 
region it is difficult to rely on the perturbative calculation, since it demands $\delta \lambda_5 >2$. } 
\label{fig:lambda5-FT}
\end{figure}

This  suggests that $\lambda_5$ extension is typically fine tuned, since it is not easy to find a perturbative $\delta \lambda_5$ for a light 
pseudo-scalar  $A$. We show this point explicitly in Fig~\ref{fig:lambda5-FT}. Most of the solutions for $\delta \lambda_5$ are already 
fine tuned, and those which are technically not fine tuned require very large values of $\delta \lambda_5$. The region with order-one values of $\delta \lambda_5$ and low fine-tuning has $m_A \lsim 1$ TeV, so the heavy Higgs bosons may be accessible at the LHC.

\section{How large can $\tan \beta$ be in natural SUSY?}
\label{sec:flavor}

\begin{figure}[h]
\begin{center}
\includegraphics[width=0.9\textwidth]{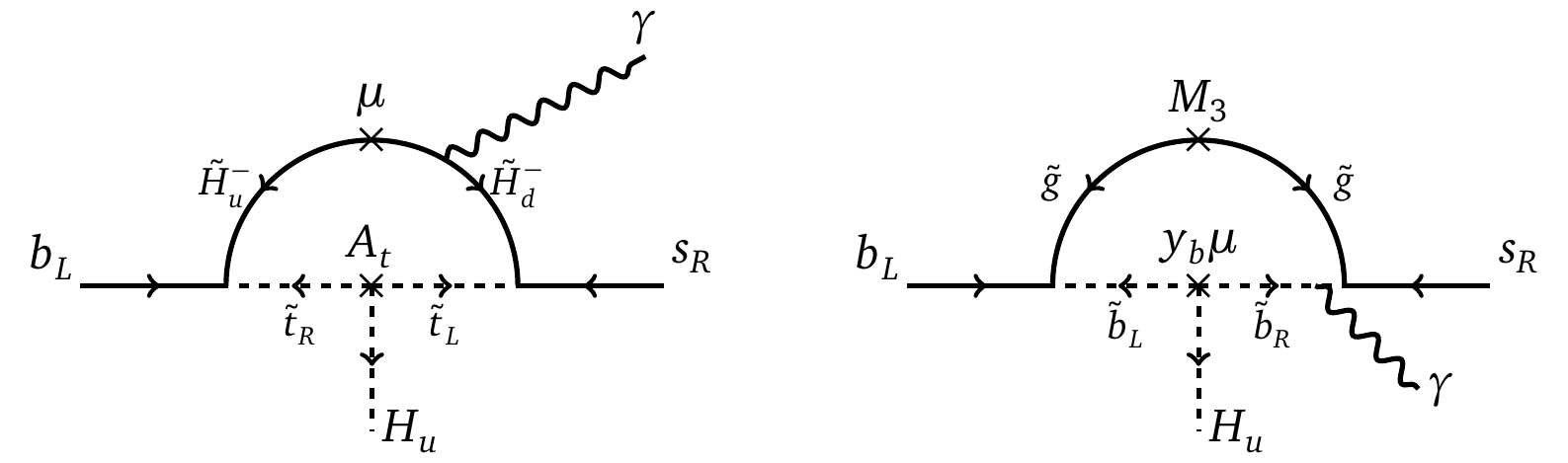}
\end{center}
\caption{Diagrams contributing to the $b \to s\gamma$ process in natural SUSY theories. The higgsino has flavor violating couplings through the CKM matrix just as the $W$ boson does, so the stop--higgsino loop at left has the same flavor factors as the SM amplitude.
} \label{fig:bsgammadiagrams}
\end{figure}

The role of $b \to s \gamma$ in natural SUSY was recently emphasized in Ref.~\cite{Ishiwata:2011ab,Blum:2012ii,Espinosa:2012in}. 
The process receives multiple contributions in supersymmetric theories that involve an insertion of the VEV of $H_u$ and thus are enhanced by a factor of $\tan \beta$ relative to the Standard Model amplitude~\cite{Oshimo:1992zd,Barbieri:1993av,Okada:1993sx,Garisto:1993jc,Baer:1997jq,Degrassi:2000qf,Carena:2000uj}. 
Two of these diagrams, one with stops and higgsinos running in the loop and one with gluinos and sbottoms, are shown in Fig.~\ref{fig:bsgammadiagrams}. (Other diagrams involve a wino or bino running in the loop; we will ignore these terms, which are small corrections in natural parts of parameter space.)\footnote{The contributions of the charged Higgs loop
is also small in most parts of the natural parameter space of $m_A$, 
if the charged Higgs is nearly degenerate with the neutral heavy Higgses. We are trying to argue that $m_A$ cannot naturally be too large, so while the charged Higgs contribution can matter at small $m_A$, it is not very relevant for our argument. Therefore we will also neglect it here. } 
From the loop diagram containing stops and higgsinos, we have a correction to the matrix element scaling like:
\beq
{\cal M}_{{\tilde t}; {\tilde h}} (b \to s\gamma) \sim m_t^2 \frac{A_t \mu}{m_{\tilde t}^4} \tan \beta.
\eeq
The measurement of the rate for $b \to s\gamma$ puts an upper bound on this correction, which we would like to interpret as an upper bound on $\tan \beta$. Such a bound would be very weak {\em if} the coefficient of $\tan \beta$ could be very small. Thus, we would like to have a lower bound on the factor $\frac{A_t \mu}{m_{\tilde t}^4}$ in front of $\tan \beta$. Fortuitously, there is an argument for each parameter that goes in the correct direction:
\begin{itemize}
\item $m_{\tilde t}$ cannot be too large because stops are needed for one-loop naturalness (canceling the top loop divergence in $m_{H_u}^2$).
\item $\mu$ cannot be too small because we have a direct constraint from LEP on the possibility of light charged particles; hence $\mu \gsim 100$ GeV~\cite{Heister:2002mn,Abdallah:2003xe,Abbiendi:2002vz,Acciarri:2000wy}. The LHC will potentially strengthen this constraint, although even raising the bound to 150 or 200 GeV will require a large luminosity at 14 TeV~\cite{Gori:2013ala,Han:2013usa,Schwaller:2013baa,Han:2014kaa}.
\item Finally, $A_t$ cannot be too small because it receives loop corrections proportional to the gluino mass $M_3$. If it takes a value much smaller than these loop corrections, this would be a new source of fine-tuning. Bounds on the gluino mass are in the vicinity of 1 TeV for a variety of scenarios, both with traditional missing momentum signatures and in cases where the gluino decays to multiple jets~\cite{Aad:2013wta,Chatrchyan:2013iqa,Chatrchyan:2014lfa,TheATLAScollaboration:2013xia,Evans:2013jna}, so it is reasonable to think that $A_t$ should not be smaller than the radiative contribution from a 1 TeV gluino.
\end{itemize}
This tells us that naturalness, used in conjunction with the measurement of $b \to s\gamma$ and experimental bounds on the gluino mass, implies an upper bound on $\tan \beta$. We should now evaluate what this bound is, numerically. The full formula is given in a convenient form in ref.~\cite{Altmannshofer:2012ks} (using the results of ref.~\cite{Freitas:2008vh}) and we will use it in the numerics, but first to get some intuition we will give some approximations that indicate how the correction depends on the soft parameters. We work in the limit $\mu^2 \ll m^2_{Q_3}, m^2_{u^c_3}$, introducing the notation $\overline{m_{\tilde t}} \equiv \left(m_{Q_3} m_{u^c_3}\right)^{1/2}$ for the geometric mean of the two stop soft masses and $r = m_{Q_3}/m_{u^c_3}$ for their ratio. Then if we assume that only the stop-higgsino loop gives a significant contribution, the general formula approximately reduces to:
\beq
\frac{{\rm Br}(B \to X_s \gamma)}{{\rm Br}(B \to X_s \gamma)_{\rm SM}} - 1 & \approx& 2.55 \tan \beta \frac{A_t \mu m_t^2}{\overline{m_{\tilde t}}^4} \left[\log \frac{\overline{m_{\tilde t}}}{\mu} \left(1+ 2.1 \frac{r^2+1}{r} \frac{\mu^2}{\overline{m_{\tilde t}}^2} \right) - 0.52 + \frac{1 + r^2}{2-2r^2} \log r \right. \nonumber \\ & & \left. -\frac{\mu^2}{\overline{m_{\tilde t}}^2} \left(0.76 \frac{3(r^2 +1)}{4r} +2.1 \frac{r^4+1}{2r(r^2-1)} \ln(r)\right)\ldots \right],
\label{eq:bsgstophiggsinoapprox}
\eeq
where omitted terms are subleading in $\tan \beta$ or in $\mu^2/\overline{m_{\tilde t}}^2$.

There are other loop corrections to $b \to s \gamma$, but they depend on masses that can naturally be heavy. The gluino loop shown at right in Fig.~\ref{fig:bsgammadiagrams} can feel flavor violation through the squark soft mass matrices; even in an MFV scenario, these need not be universal, because---for example---$m_Q^2$ can contain a piece proportional to $V^\dagger y_u^2 V$ where $y_u^2$ is a diagonal matrix of up-type Yukawas~\cite{Altmannshofer:2012ks}. However, the gluino loop  involves the right-handed sbottom, which need not be light for naturalness. In fact, in some natural SUSY scenarios it must be heavy to avoid FCNCs~\cite{Brust:2011tb, Mescia:2012fg}. Even if we assume that the right-handed sbottom mass is near the left-handed sbottom and stop masses, we find that the gluino loop is usually subdominant to the stop--chargino loop for natural parameter values. The wino loop is suppressed by a smaller coupling as well as potentially the heaviness of the wino mass. Thus, it is reasonable for us to focus on the stop--chargino loop. In principle, other loop corrections {\em could} cancel it, but this is in itself a tuning.

\subsection{Natural choices for $A_t$}
 
The simplest estimate for the smallest natural choice of $A_t$, assuming running from a relatively low scale $\Lambda$, is
\beq
A_t^{\rm loop} \approx -\frac{2}{3 \pi^2} g_3^2 M_3 \log\frac{\Lambda}{M_3} \approx -230~{\rm GeV} \left(\frac{M_3}{1~{\rm TeV}}\right) \log_{10} \frac{\Lambda}{M_3}.
\label{eq:AtoneloopRG}
\eeq
If we run from a higher scale, we can do a somewhat more careful estimate by resumming large logarithms.

\begin{figure}[!h]
\begin{center}
\includegraphics[width=0.4\textwidth]{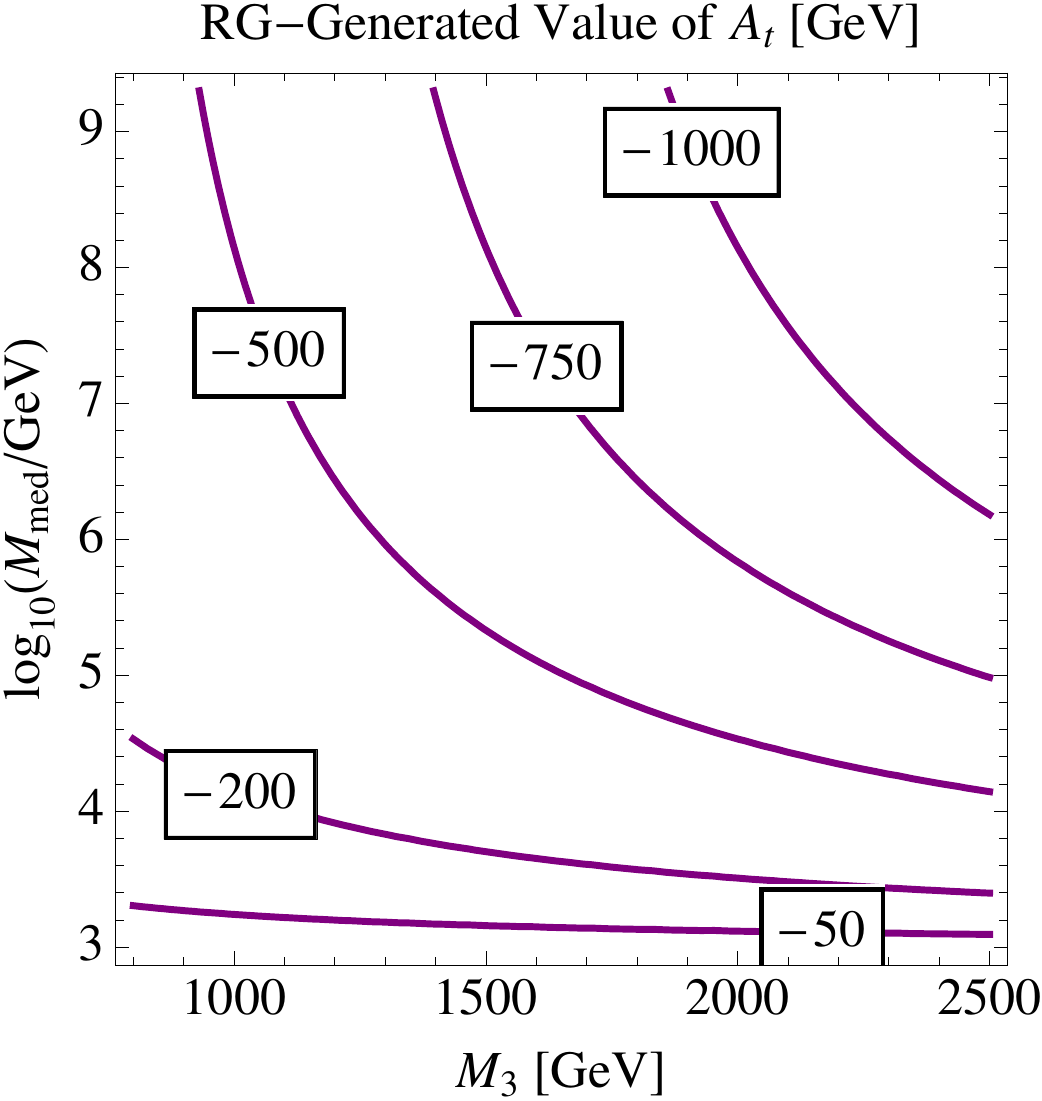}
\end{center}
\caption{The low-scale value of $A_t$ generated from solving the RGE with $A_t = 0$ at a scale $M_{\rm med}$ and a low-scale gluino mass $M_3$.}
\label{fig:AtfromRG}
\end{figure}

What we have called $A_t$ is really $a_t/y_t$, where $a_t$ is the coefficient of the three-scalar operator in the Lagrangian. Keeping only the one-loop terms involving $g_3$ or $y_t$, the RG evolution of $a_t$ is related to that of the gluino mass by the equation (e.g.~\cite{Martin:1997ns})
\beq
\frac{d}{d\log\mu} a_t & = & \frac{1}{16 \pi^2} \left(\left(18 y_t^2 - \frac{16}{3} g_3^2\right) a_t + \frac{32}{3} y_t g_3^2 M_3\right). 
\eeq
If we assume that $a_t \approx 0$ at some mediation scale $M_{\rm med}$ (as is true in a number of models, including gauge mediation), we can use this equation together with the RGEs for $g_3, y_t,$ and $M_3$ to plot the low-scale value of $A_t$ as a function of the low-scale gluino mass parameter $M_3$ and the mediation scale. We show this in Fig.~\ref{fig:AtfromRG}.

What we learn from this is that typically, the RG contribution to $A_t$ ranges from $-200~{\rm GeV}$ to $-750~{\rm GeV}$ over a wide range of mediation scales and for gluinos near 1 TeV. Thus, a smaller trilinear coupling $A_t$ will generally imply some tuning of a positive tree-level value at the mediation scale against a negative loop correction from gluinos. (Here we use ``tree-level'' loosely for the value of $A_t$ at the mediation scale; in a given model, it may arise from loops, but we distinguish it from the contribution generated in the RGE.) We will quantify this tuning in an intuitive way. Given that $A_t$ is a sum $A_t = A_t^{\rm tree} + A_t^{\rm loop}$, we can measure a tuning by the amount of cancelation:
\beq
\Delta_{A_t} \equiv \frac{\left|A_t^{\rm tree}\right| + \left|A_t^{\rm loop}\right|}{\left|A_t^{\rm tree} + A_t^{\rm loop}\right|}. 
\label{eq:Attuning}
\eeq
In the regime where the two terms nearly cancel, this behaves similarly to other tuning measures like that of Barbieri and Giudice~\cite{Barbieri:1987fn}. If there is no significant cancelation (e.g. if $A_t$ at the mediation scale is much larger than the gluino-generated term), it asymptotes to $1$. This is a desirable property for a tuning measure to have, because we would like to be able to compute a combined tuning in multiple variables as a product of independent tunings in each variable.

\subsection{The uplifted Higgs region}

\begin{figure}[h]
\begin{center}
\includegraphics[width=0.8\textwidth]{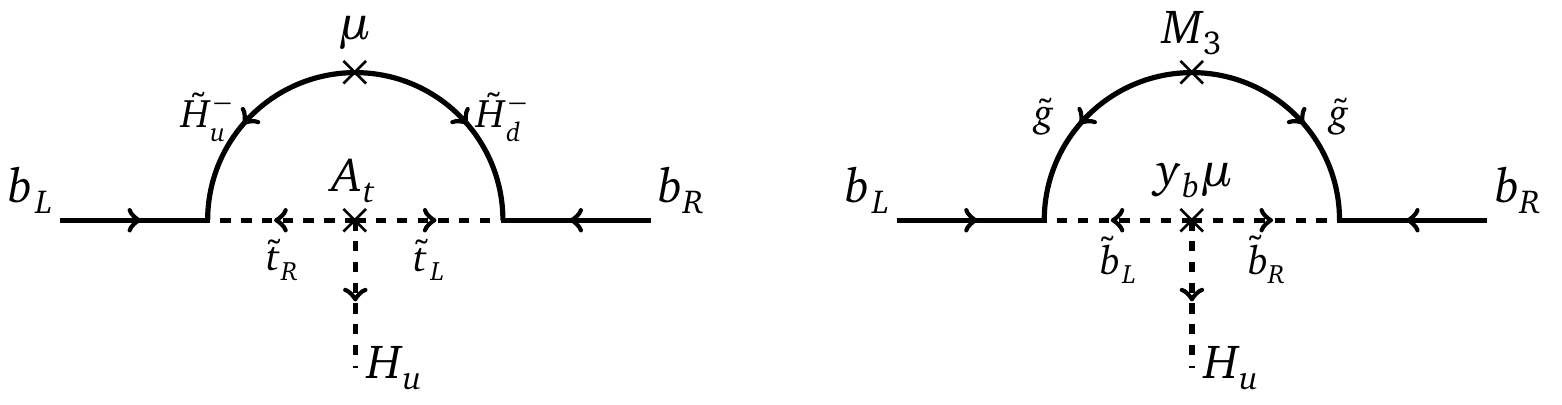}
\end{center}
\caption{Diagrams contributing to the ``wrong-Higgs'' Yukawa coupling $H_u^\dagger Q d^c$, which can be a dominant contribution to the $b$-quark mass for very large $\tan \beta$.
} \label{fig:wronghiggsyukawa}
\end{figure}

As $\tan \beta$ increases, the Yukawa couplings needed to generate the $b$ and $\tau$ masses from the VEV of $H_d$ become large. However, a new source of masses arises from loop effects that generate the ``wrong-Higgs'' Yukawa couplings $H_u^\dagger Q d^c$ and $H_u^\dagger L e^c$. For sufficiently large $\tan \beta$ we can think of the $b$ and $\tau$ masses as arising entirely for these effects, in what has been called the uplifted supersymmetric Higgs region of parameter space~\cite{Dobrescu:2010mk,Altmannshofer:2010zt}. In this part of parameter space, we must exercise some caution in our argument about the size of the $b \to s\gamma$ amplitude. The same loop diagram that generates the wrong-Higgs bottom quark Yukawa coupling also generates $b \to s\gamma$, when one external $b$ quark is replaced by a strange quark and a photon is attached to an internal line. (Compare the loops generating Yukawas in Fig.~\ref{fig:wronghiggsyukawa} to those for $b \to s\gamma$ in Fig.~\ref{fig:bsgammadiagrams}.) As a result, $b \to s\gamma$ is no longer enhanced by a factor of $\tan\beta$ relative to the $b$-quark mass, and we should be concerned that data on $b \to s\gamma$ can't actually rule out very large values of $\tan \beta$.

This concern is conceptually reasonable but proves to be numerically unfounded. The uplifted region of parameter space lies at very large values of $\tan \beta$ and also requires large values of $\mu$, putting it outside of what we consider to be natural SUSY parameter space. In all of our computations, we will use the formulas of Ref.~\cite{Altmannshofer:2012ks}, in which the corrections to the $b \to s\gamma$ amplitude are proportional to $\tan \beta/(1 + \epsilon_b \tan \beta)$, where $\epsilon_b$ in the denominator is a loop factor correcting for the wrong-Higgs contribution to the $b$-quark mass. The statement that the uplifted regime does not change our conclusion is that $\epsilon_b \tan \beta$ is at most ${\cal O}(1)$ for reasonable input parameters, whereas removing the bound at large $\tan \beta$ would require that it be $\gg 1$.

It is easy to see that naturalness is in tension with the uplifted regime by inspection of the loop corrections. The approximate result for $\epsilon_b$ in the limit $m_{Q_3}^2 = m_{u^c_3}^2 = m_{d^c_3}^2 \equiv m_{\tilde q}^2$ assuming $M_3^2 \gg m_{\tilde q}^2 \gg \mu^2$ is
\beq
\epsilon_b \approx \frac{1}{16 \pi^2} \left\{\frac{8 g_s^2}{3} \frac{\mu}{M_3} \left[\log\frac{M_3^2}{m_{\tilde q}^2} \left(1 + 2 \frac{m_{\tilde q}^2}{M_3^2}\right) - 1\right] + \frac{y_t^2 s_\beta^2 A_t \mu}{m_{\tilde q}^2} \left(1 - \frac{\mu^2}{m_{\tilde q}^2} \log \frac{m_{\tilde q}^2}{\mu^2}\right) + \ldots \right\}.
\eeq
Numerically, we expect the gluino loop contribution to $\epsilon_b$, which is 
$\sim \mu/M_3$, to dominate in most of the natural SUSY parameter space. Note that for naturalness, we prefer $\mu$ as small as possible (close to 100 GeV), whereas experimentally we know that $M_3 \gsim 1~{\rm TeV}$. Furthermore, the gluino loop drags the stop and sbottom soft masses up, so the log is rarely large. Estimating $\mu/M_3 \lsim 0.2$ and $\log(M_3^2/m_{\tilde q}^2) \lsim 3$, we see that $\epsilon_b \lsim 10^{-2}$, so that $\epsilon_b \tan \beta$ becomes an order-one number only at $\tan \beta \sim 100$. One could try to get around this conclusion by choosing very large values of $A_t$ to enhance the second term, but this is not very well-motivated and potentially runs into problems with vacuum stability~\cite{Kusenko:1996jn}. Increasing the first term requires going to large $\mu$ and thus indicates significant tree-level tuning for electroweak symmetry breaking. In short, the uplifted regime is of little relevance for a study of natural SUSY, and will not interfere with our inference of a bound on $\tan \beta$ from $b \to s\gamma$ and naturalness.

\subsection{Interpreting the experimental results on $b \to s \gamma$}
\label{sec:bsgammaboundresults}

For the experimental bound on $b \to s\gamma$, we will follow Ref.~\cite{Altmannshofer:2012ks} in taking the SM prediction to be~\cite{Misiak:2006zs} ${\rm Br}(B \to X_s \gamma)_{\rm SM} = \left(3.15 \pm 0.23\right) \times 10^{-4}$ and the experimental value to be~\cite{Lees:2012ufa,Amhis:2012bh} ${\rm Br}(B \to X_s \gamma)_{\rm exp} = \left(3.43 \pm 0.22\right) \times 10^{-4}$. Given these values, we estimate that at 95\% confidence level the ratio $R_{b s\gamma}$ of the true value of the branching ratio to its Standard Model value lies in the range
\beq
0.90 \leq R_{bs\gamma} \leq 1.32.
\eeq
Because the data prefer a slightly high value relative to the Standard Model, constraints are weaker on the scenario where new physics constructively interferes with the SM amplitude. This happens when $\mu A_t > 0$. Because small $A_t$ is easiest to achieve if the RG contribution from the gluino dominates, this corresponds to a negative sign for $\mu M_3$. The case $\mu A_t < 0$, arising if $\mu$ and the gluino mass term have the same sign, is more strongly constrained.

\begin{figure}[!h]
\begin{center}
\includegraphics[width=0.75\textwidth]{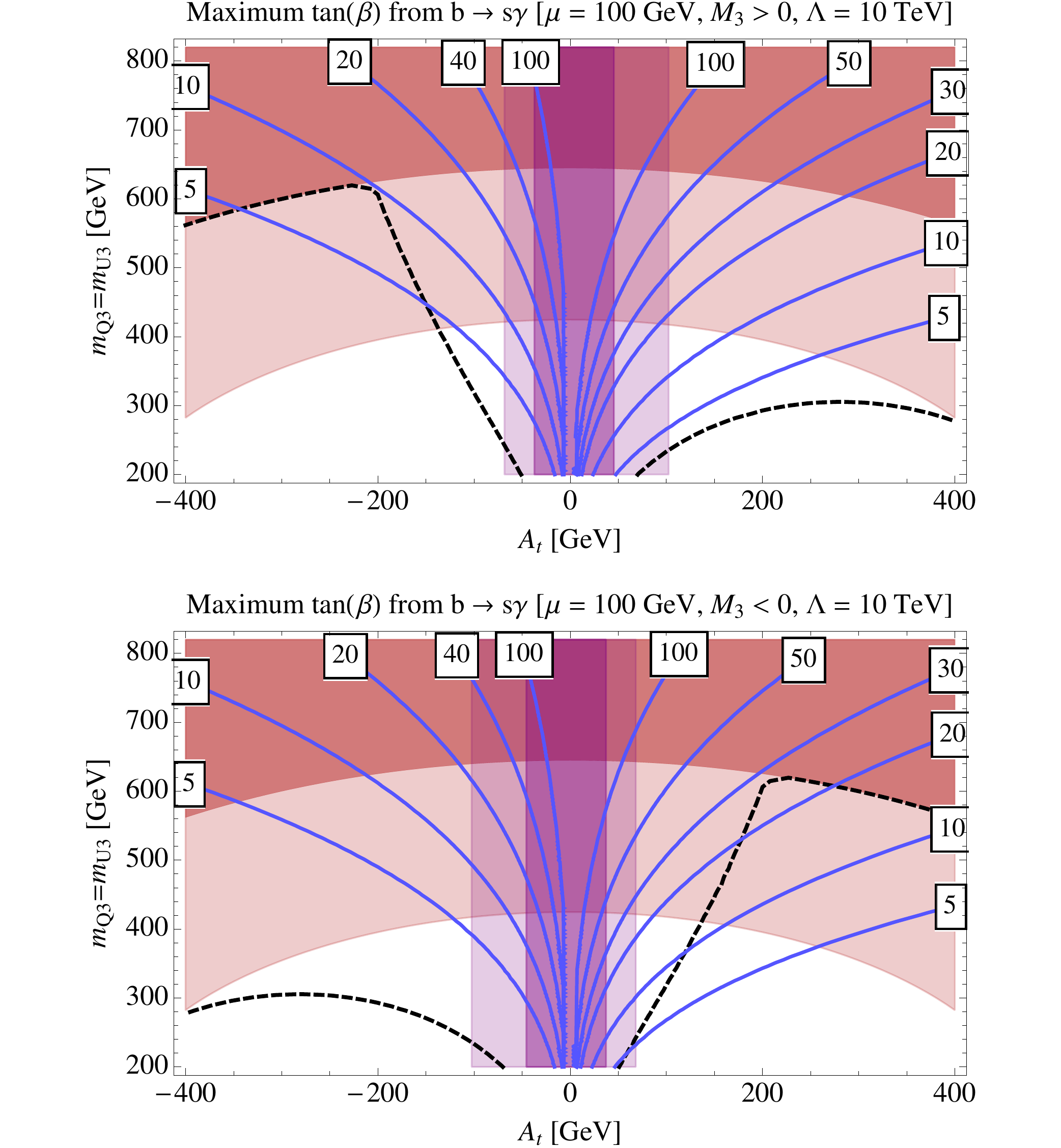}
\end{center}
\caption{Constraints arising from $b \to s \gamma$. Here we have fixed $\mu = 100$ GeV (and $|M_3| = 1.3$ TeV) and plot blue solid lines for contours of the largest allowed $\tan \beta$ as a function of the stop mixing parameter $A_t$ and the stop soft mass parameter. The shaded regions are disfavored by naturalness: the purple regions at small $A_t$ involve tuning $\Delta_{A_t} = 5$ (lighter region) and 10 (darker region). The red shaded regions correspond to $\Delta_{\tilde t} = 5$ (lighter) and 10 (darker) tuning in $m_{H_u}^2$ from the stop loop contribution. The region above the black dashed lines has combined tuning $\Delta > 10$. The plots with different signs of $M_3$ have different tuning measures because the loop-generated $A_t$ always has the opposite sign to $M_3$.}
\label{fig:bsgammaconstraints}
\end{figure}

We have plotted the largest allowed value of $\tan \beta$, with various naturalness constraints superimposed, in Fig.~\ref{fig:bsgammaconstraints}. In this figure $\mu$ is fixed to 100 GeV. We have also fixed $\left|M_3\right| = 1.3$ TeV, $m_{d_3} = 2$ TeV, $M_2 = 0.5$ TeV, and $\zeta$ (a parameter defined in ref.~\cite{Altmannshofer:2012ks} related to the relative size of various MFV terms in the soft mass matrices) equal to 0.5. We assume that running begins at $\Lambda = 10$ TeV, a fairly extreme limit of low-scale SUSY breaking, in order to be conservative about tuning measures. Bounds on $\tan \beta$ become stronger if $\mu$ increases. The plot is relatively insensitive to the other parameters, but we have included them for concreteness. We have checked that including the stop--chargino loop alone, with all other superpartners decoupled, makes very little difference in the result. We plot two cases with two different signs of $M_3$ (relative to $\mu$). The sign of $M_3$ determines the sign of $A_t^{\rm loop}$ which enters in the tuning measure Eq.~\eqref{eq:Attuning}.

\begin{figure}[!h]
\begin{center}
\includegraphics[width=0.75\textwidth]{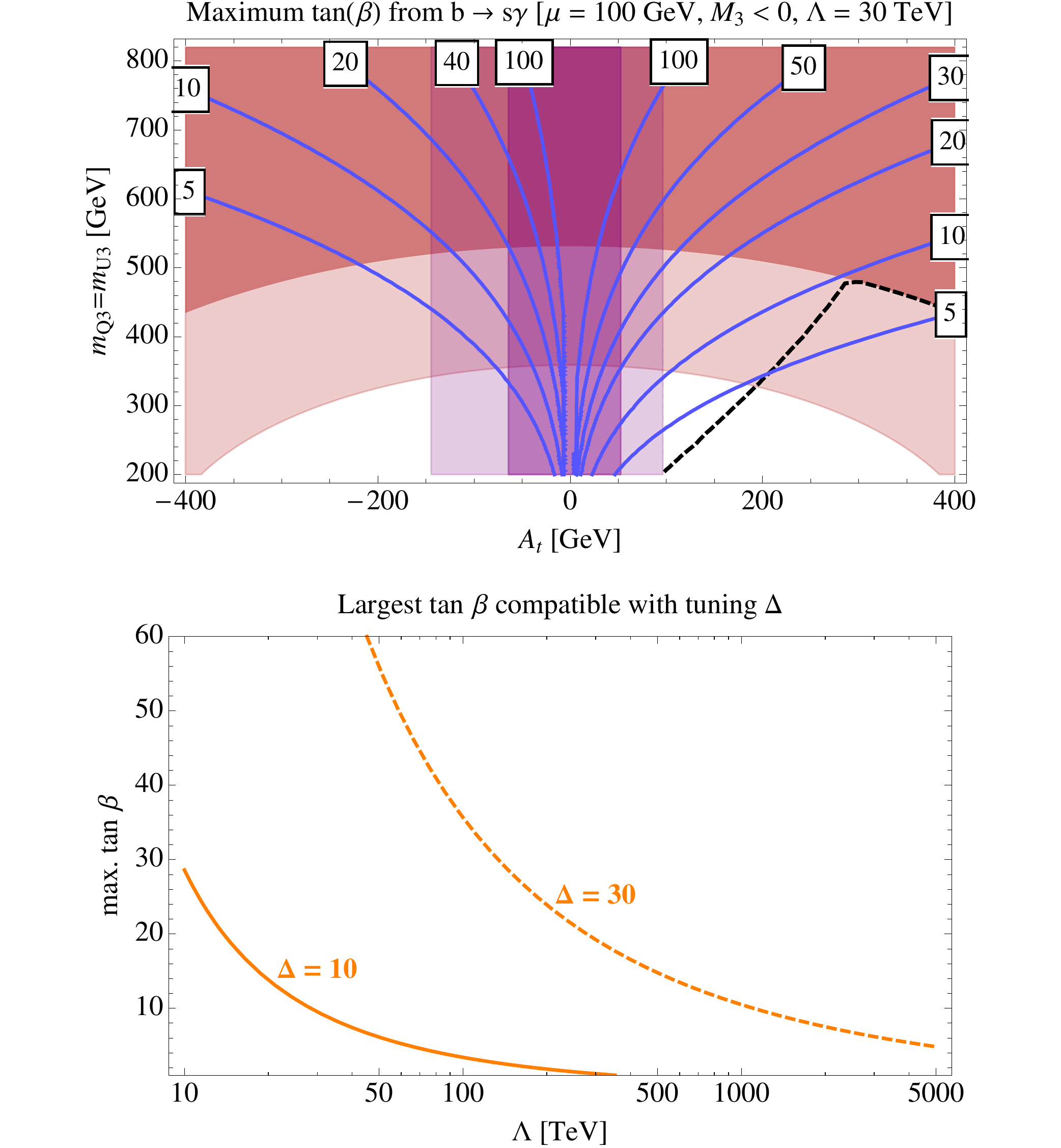}
\end{center}
\caption{Constraints arising from $b \to s \gamma$. The upper plot is just like the lower panel of Fig.~\ref{fig:bsgammaconstraints}, except that we set the supersymmetry mediation scale $\Lambda$ to 30 TeV instead of 10 TeV. The extra running means that increased tuning is required: both $\Delta_{A_t}$ and $\Delta_{\tilde t}$ are larger. As a result, requiring $\Delta < 10$ now imposes a stronger constraint, $\tan \beta < 9.5$. In the lower plot we show how this constraint evolves with the mediation scale, allowing for stop-sector tuning by a factor of either 10 (solid orange line) or 30 (dashed orange line). Already for a 100 TeV mediation scale the constraint is $\tan \beta < 3.4$ if we require $\Delta < 10$.}
\label{fig:bsgammavscutoff}
\end{figure}

The naturalness constraints are defined in terms of two tunings. First, large stop soft masses correspond to a tuning of the up-type Higgs soft mass parameter, which is quantified by~\cite{Kitano:2006gv,Perelstein:2007nx}
\beq
\Delta_{\tilde t} = \left|\frac{3 y_t^2}{4\pi^2} \frac{m^2_{Q_3} + m^2_{u_3} + A_t^2}{m_h^2} \log \frac{\Lambda}{\overline{m_{\tilde t}}}\right|.
\label{eq:stoptuning}
\eeq
The second tuning arises for small values of $A_t$, as quantified in the expression $\Delta_{A_t}$ of Eq.~\eqref{eq:Attuning}. In Fig.~\ref{fig:bsgammaconstraints}, regions of large $\Delta_{\tilde t}$ are shaded red and regions of large $\Delta_{A_t}$ are shaded purple. One can see that large values of $\tan \beta$ are allowed only if $A_t$ is small or the stop masses are large, indicating that at least one of these tuning measures is becoming large. For instance, there are two corners of parameter space where $\Delta_{A_t} = \Delta_{\tilde t} = 5$, one at negative $A_t$ and one at positive $A_t$ (where the sign is understood relative to that of $\mu$). In the case $M_3 < 0$, at the former point, the largest 95\% CL allowed value of $\tan \beta$ is less than 10; at the latter, it is about 25. Thus, as noted above, the case of positive $A_t$ is less strongly constrained.

We have no particular reason to think that cancellations in $m_{H_u}^2$ and in $A_t$ will happen at the same point in parameter space, although perhaps one could imagine a model in which this is true. If the tunings {\em are} independent, we can think of an overall tuning $\Delta = \Delta_{\tilde t} \Delta_{A_t}$ which is simply the product of the two individual tunings. In other words, if we have to adjust two unrelated parameters to the 10\% level, this may reasonably be thought of as a 1\% tuning in parameter space. With such independent tunings in mind, we have plotted dashed black contours in Fig.~\ref{fig:bsgammaconstraints} that show where $\Delta = 10$. We see that the combined tuning is mildest whenever $M_3 A_t < 0$, which is driven by the fact that $\Delta_{A_t}$ prefers $A_t$ to be either near its loop-generated value or much bigger.

The most optimistic region of parameter space has $\mu A_t > 0$ (so that the new physics contribution constructively interferes with the SM and improves agreement with data) and $A_t M_3 < 0$ (so that the trilinear can be mostly generated from the RG). From the figure, we can see that this marginally allows $\tan \beta \approx 28$ with a combined tuning $\Delta \approx 10$ coming almost entirely from the stop mass $m_{\tilde t} \approx 600~{\rm GeV}$. The plots make it clear that allowing $\tan \beta > 30$ will require either quite heavy stops---out of the range that can be considered truly natural---or a cancelation in $A_t$, or both. We think that it is very conservative to conclude that {\em generic} natural SUSY requires $\tan \beta < 30$.

In fact, we are usually understating the required cancelation in $A_t$, because most reasonable models will run from a higher UV scale and generate values of $A_t$ a factor of 2 or more larger than we have considered. Even a slightly larger amount of running produces a significantly stronger conclusion, as we illustrate in Fig.~\ref{fig:bsgammavscutoff}. Beginning the RGE at 30 TeV instead of 10 TeV produces a larger value of $A_t^{\rm loop}$ and also increases the stop-generated contribution to $m_{H_u}^2$. In this case, the conclusion is already that $\tan \beta < 10$. We show how the bound on $\tan \beta$ changes with the mediation scale in the lower panel of Fig.~\ref{fig:bsgammavscutoff}. Running from 100 TeV already requires $\tan \beta < 3.4$ for consistency with an overall tuning $\Delta < 10$. (In fact, the bound already hits $\tan \beta = 1$ when the mediation scale $\Lambda \approx 350$ TeV, indicating that models of high-scale SUSY breaking will require significant fine-tuning for compatibility with the $b \to s \gamma$ measurement.) Although the choice of a tuning measure is to some extent a matter of taste, it is clear that accommodating $\tan \beta \gsim 10$ requires both {\em very} low-scale mediation and a mild tuning. We also show, with the dashed orange line in the lower panel of Fig.~\ref{fig:bsgammavscutoff}, that allowing for more tuning significantly increases the range of allowed $\tan \beta$. If we allow $\Delta = 30$ rather than 10, we can accommodate $\tan \beta = 30$ even with running from 100 TeV, and $\tan \beta = 10$ even with running from 1000 TeV. Still, high-scale SUSY breaking is highly constrained even allowing for this larger amount of tuning.

As a simpler estimate, we can use the one loop RG approximation eq.~\eqref{eq:AtoneloopRG} for $A_t^{\rm loop}$, write the average stop mass in terms of the tuning $\Delta_{\tilde t}$ from eq.~\eqref{eq:stoptuning}, and use the leading term in equation~\eqref{eq:bsgstophiggsinoapprox} to estimate that
\beq
2.55 \tan \beta \frac{A_t \mu m_t^2}{\overline{m_{\tilde t}}^4} \log\frac{\overline{m_{\tilde t}}}{\mu} 
\lsim 0.32,
\eeq
implying
\beq
\tan \beta \lsim 28 \left(\frac{\Delta_{\tilde t}}{10}\right)^2 \left(\frac{100~{\rm GeV}}{\mu}\right) \left(\frac{1.3~{\rm TeV}}{\left|M_3\right|}\right) \frac{2}{\log\frac{\overline{m_{\tilde t}}}{\mu}} \frac{2}{\log\frac{\Lambda}{|M_3|}}  \left(\frac{2}{\log\frac{\Lambda}{\overline{m_{\tilde t}}}}\right)^2.
\label{eq:tanbetabound}
\eeq
This is a useful check that the more detailed numerical results are reasonable. The $\left(\log \Lambda\right)^{-3}$ behavior explains the rapid improvement of the bound as we increase $\Lambda$ above 10 TeV that we saw in Fig.~\ref{fig:bsgammavscutoff}.

\subsection{Comment on  $B_s \to \mu^+ \mu^-$.}

This very rare process is often quoted as the best possible constraint on SUSY with large $\tan \beta$. Indeed the most important 
SUSY contribution to $B_s \to \mu^+ \mu^- $ is proportional to $\tan^3 \beta$~\cite{Babu:1999hn,Isidori:2001fv}, 
and therefore is naively expected to be very 
sensitive to natural SUSY.  However we find that all the constraints that we get from this process are subdominant to $b \to s \gamma $
constraints. There is a simple explanation for why this happens. Although the matrix element is enhanced by $\tan^3 \beta $, it is also suppressed
by $m_A^2$. As we have learned in Sec.~\ref{sec:treelevelFT}, in the large $\tan \beta $ limit the fine tuning of EWSB stays approximately 
constant along the contours of $m_A \tan \beta = const$. Therefore, effectively the matrix element is enhanced 
only by a single power of 
$\tan \beta$, precisely as is $b \to s \gamma$.

\begin{figure}[ht]
\begin{center}
\includegraphics[width=.8\textwidth]{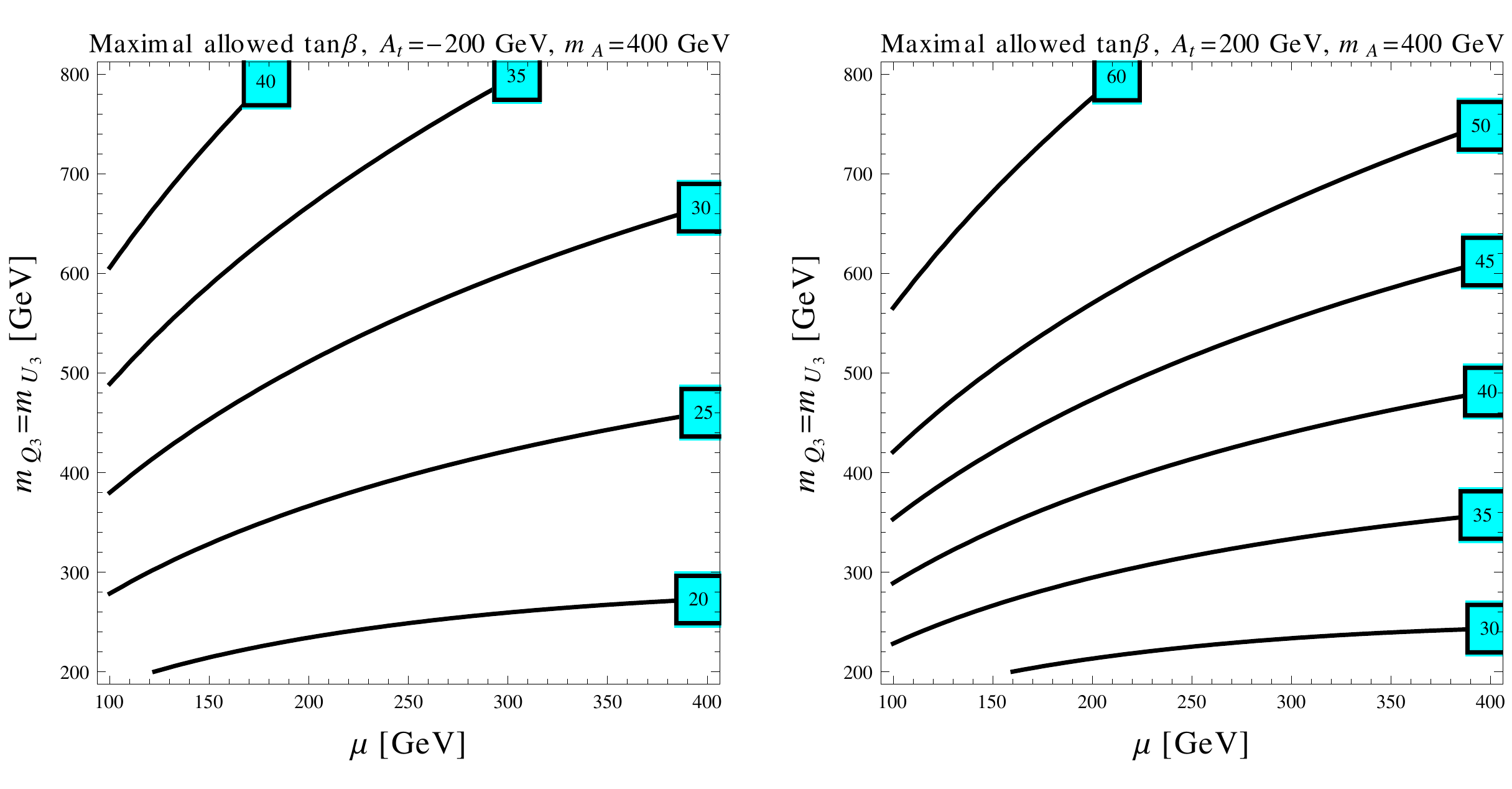}
\end{center}
\caption{Constraints arising from $B_s \to \mu^+ \mu^-$ for $m_A = 400$~GeV.}
\label{fig:bs400}
\end{figure}

On the other hand, the rate of $B_s\to \mu^+ \mu^- $ is measured to much worse precision than $b \to s \gamma$. 
While the process $b \to s \gamma $ is measured to the precision of better than 10\%,  Ref.~\cite{Aaij:2012nna} gives the following 95\%
CL bound on $B_s \to \mu^+ \mu^-$:
\beq 
1.1 \times 10^{-9} < BR(B_s \to \mu^+ \mu^-)_{exp} < 6.4\times 10^{-9}
\eeq
Based on the SM prediction~\cite{Buras:2012ru} 
\beq
BR(B_s \to \mu^+ \mu^-)_{SM} = (3.32\pm 0.17)\times 10^{-9},
\eeq
from these equations we estimate that at 95\% confidence level
\beq
0.31 < R_{B_s\to \mu^+ \mu^-} < 1.95.
\eeq
The lower bound is meaningless in large $\tan \beta$ regime: in SUSY one cannot get $R_{B_s\to \mu^+ \mu^-}$ 
smaller than 0.5, unless the SUSY contribution is dominated by the $Z$-penguin.\footnote{The reason for this is that there is a contribution with $H^0$ exchange that interferes destructively with the Standard Model, and a contribution with $A^0$ exchange that does not interfere and is equal to the $H^0$ amplitude to the extent that $m_A \approx m_{H^0}$. Thus the squared matrix element goes as $\left|A_{\rm SM} - A_{\rm NP}\right|^2 + \left|A_{\rm NP}\right|^2 \geq \frac{1}{2} \left|A_{\rm SM}\right|^2$. If the $Z$ penguin contribution matters, this argument is no longer strictly true. But the $Z$ penguin goes only as $\tan^2 \beta$ and is suppressed by the mass insertion $\delta_{LR}$ in the up sector (see~\cite{Chankowski:2000ng} for 
relevant expressions), so is generally expected to be less important.}
On the other hand the 
upper bound is weak, allowing ${\cal O}(1)$ deviations from the SM-predicted values. Therefore the bounds on $\tan \beta $ one gets 
from this process are much weaker than those one gets from $b \to s \gamma$. To illustrate these points we show these bounds, considering
(as in the $b \to s \gamma $ example) only the higgsino loop contribution, in Fig.~\ref{fig:bs400}, showing the maximal allowed values of 
$\tan \beta $ for $m_A = 400$~GeV. We see that these constraints are very clearly subdominant to the $b \to s \gamma$ constraints 
in Fig.~\ref{fig:bsgammaconstraints}. At higher $m_A $ these constraints quickly decouple.  

\subsection{The Dirac loophole}

If the gluino has a {\em Dirac mass} rather than the standard Majorana mass, our argument breaks down because $A_t$ can naturally be smaller, protected by $R$-symmetry. Supersymmetry with Dirac gauginos has received a great deal of recent attention: see (for example) refs.~\cite{Fox:2002bu,Nelson:2002ca,Kribs:2007ac,Davies:2011mp,Frugiuele:2011mh,Brust:2011tb,Kribs:2012gx,Benakli:2012cy,Han:2012cu,Arvanitaki:2013yja,Csaki:2013fla,Kribs:2013eua,Bertuzzo:2014bwa}. An $R$-symmetry forbids a $\mu$-term as the higgsino mass, so these models typically involve new doublets that pair with the usual Higgs doublets to form massive higgsinos. The $A$-term is also forbidden. Depending on how and to what extent the $R$-symmetry is broken, a remnant of our argument may survive in these models. For the most part, however, we expect that these models evade our argument and that a detailed look at the EWSB conditions and naturalness in these theories will require a completely different perspective. Some aspects of naturalness in such theories have been addressed in refs.~\cite{Arvanitaki:2013yja,Csaki:2013fla,Bertuzzo:2014bwa}. Although this loophole exists, models with Dirac gauginos are necessarily more baroque than traditional SUSY models, and we do not feel that they undermine the motivation for viewing heavy Higgs bosons as key channels in which to search for naturalness.

\section{Outlook}

The traditional harbingers of SUSY naturalness are higgsinos at tree level, stops at one loop, and gluinos at two loops. Higgsinos, being produced only through the electroweak interactions, are difficult to constrain at hadron colliders. Stops and, especially, gluinos are easier to search for directly due to their large QCD cross sections. But if $R$-parity is violated, the spectrum is compressed, or decays go through a hidden sector, traditional missing momentum searches for stops and gluinos can be dramatically weakened. Optimized searches for these ``hidden SUSY'' cases are receiving increased attention. One of our goals in this paper is to argue that searches for heavy Higgs bosons provide another way to address such scenarios.

Heavy Higgs bosons, unlike superpartners, have predictable decays to pairs of Standard Model particles. The neutral boson $H^0$ will decay to $\tau$'s and $b$'s at large $\tan \beta$, and to tops, light Higgs bosons, $Z$ bosons and $W$ bosons at smaller $\tan \beta$. The $ZZ$ ``golden channel'' is one interesting search mode, and its rate is linked to the $hh$ channel by the Goldstone equivalence theorem, as explained in Appendix~\ref{sec:branchingratiocomment}. Although extensions of the MSSM might open new decay modes of the heavy Higgses, it seems unlikely that these decays dominate, especially given the SM-like nature of the light Higgs as measured so far. Thus, heavy Higgs searches offer a window on naturalness that is less easily dodged by clever model-building than other SUSY searches.

One recent survey of the reach of LHC Run II for heavy Higgs bosons is Ref.~\cite{Arbey:2013jla}, which shows that the $H \to \tau^+ \tau^-$ channel can reach above 1 TeV for large $\tan\beta$ while $H \to t{\bar t}$ can reach above 1 TeV at small $\tan \beta$. The intermediate $\tan \beta$ regime is more difficult to probe and could deserve increased effort, given the added motivation that arises when viewing these searches as an additional probe of naturalness. Other recent theoretical work on signals of heavy Higgs bosons includes refs.~\cite{Brownson:2013lka,Carena:2013qia,Craig:2013hca,Craig:2012pu,Eberhardt:2013uba,Dumont:2014wha}.

It is interesting to ask to what extent our naturalness bounds on heavy Higgs masses can be improved in the future. We do not expect significant theoretical improvements in the Standard Model prediction of $b \to s\gamma$ in the future, due to irreducible uncertainties~\cite{Lee:2006wn,Benzke:2010js}. The currently less constraining measurement of $B_s \to \mu^+ \mu^-$ might play a more interesting role in the future. The LHCb result~\cite{Aaij:2012nna} is dominated by statistical uncertainties. Future data is expected to improve the error bar to 10\% precision~\cite{Bediaga:2012py}. With such an improved measurement, the constraints from $B_s \to \mu^+ \mu^-$ would become an important supplement to the $b \to s\gamma$ bound in naturalness arguments regarding the large $\tan \beta$ region.

Another way that our arguments could become somewhat stronger in the future is through an improved lower bound on the higgsino mass parameter $\mu$, which will in turn require smaller values of $\tan \beta$ to accommodate the same constraint on $b \to s\gamma$. However, higgsino searches are difficult. A more promising route to a stronger bound is through improved bounds on gluino masses, since these feed into $A_t$ at loop level. Although gluino signals are susceptible to being ``hidden'' in various ways, they are less so than stops, and in fact bounds on gluinos exist even with complicated decay chains lacking missing energy~\cite{Evans:2013jna}. These bounds should improve early in Run 2 of the LHC, which will allow a stronger statement to be made about the heavy Higgs masses expected from naturalness arguments.

We emphasize that heavy Higgs boson searches provide a robust way to constrain natural models of supersymmetry. Although they are already a part of the LHC's suite of new physics searches, we believe that they should be viewed as part of the naturalness program and accorded a correspondingly intense focus.

\section*{Acknowledgments}

We thank Michelangelo Mangano and Carlos Wagner for interesting discussions. We thank an anonymous referee for interesting comments about $B_s \to \mu^+ \mu^-$ and its possible future role. MR thanks Joao Guimaraes da Costa, Tomo Lazovich, Masahiro Morii, and Hugh Skottowe for the discussion that prompted appendix~\ref{sec:branchingratiocomment}. AK is supported by the Center for the Fundamental Laws of Nature at Harvard. AS is supported by the National Science Foundation Award ID 1067976.

\appendix

\section{$H \to hh$, $H \to ZZ$, and Goldstone equivalence: a comment on branching ratios}
\label{sec:branchingratiocomment}

One interesting search channel for a heavy Higgs is $H \to hh$, which is particularly appealing since the Standard Model rate for events with two Higgs bosons is very small~\cite{Dolan:2012ac,Craig:2013hca,Gupta:2013zza,Ellwanger:2013ova,Barr:2013tda,Dolan:2013rja,No:2013wsa,Baglio:2014nea,Chen:2014xra}. On the other hand, the dominant Higgs decay is to $b{\bar b}$, a challenging signal to pull out of background. Given that the very clean $h \to ZZ^* \to 4 \ell$ channel played a key role in the discovery of the 125 GeV Higgs boson, it is interesting to ask when, and to what extent, the $H \to hh$ decay mode dominates over $H \to ZZ$. Answers to this question may be extracted from the literature, but are often expressed in rather technical forms. For example, in ref.~\cite{Craig:2013hca}, we learn that the coupling $g_{Hhh}$ is proportional to $\left(3 m_A^2 - 2 m_h^2 - m_H^2\right)\left(\cos(2\beta-2\alpha) - \cot(2\beta) \sin(2\beta-2\alpha)\right)-m_A^2$. Even an MSSM aficionado might have to resort to numerical estimates to have much intuition for what such an expression means. On the other hand, numerically, one can see from plots (e.g. in refs.~\cite{Arbey:2013jla} or~\cite{Craig:2013hca}) that $\Gamma(H \to hh)$ is typically about an order of magnitude larger than $\Gamma(H \to ZZ)$.

In fact, in most models it will be true that $\Gamma(H \to hh) \approx 9~\Gamma(H \to ZZ)$, which follows straightforwardly from the Goldstone boson equivalence theorem. Corrections are expected to be of order $m_h^2/m_H^2$. This result is likely known to experts but we have not seen it in the literature, so we will explain it here. It offers a useful rule-of-thumb for experimentalists considering whether to undertake a search for Higgs pair production. Assuming this factor of 9 between the heavy Higgs branching ratios, one can ask whether a planned search for Higgs pair production can beat the cleaner, but rarer, $ZZ \to 4 \ell$ signal.

The factor of 9 in the rate comes from a combinatoric factor of 3 in the amplitude that we can explain using a strategy that has appeared in ref.~\cite{Gupta:2012fy}, namely working in the basis of VEV eigenstates. We will denote by $h$ the linear combination of fields that has a VEV, and $H$ the orthogonal combination:
\beq
h = \sin \beta~H_u + \cos \beta~H_d^\dagger & = & \begin{pmatrix} i G^+\\ (v + h^0 + i G^0)/\sqrt{2}\end{pmatrix}, \\
H = -\cos \beta~H_u + \sin \beta~H_d^\dagger & = & \begin{pmatrix} i H^+ \\ (H^0 + i A^0)/\sqrt{2}\end{pmatrix}.
\eeq 
Notice that we are working not just with the real components of the Higgs fields but with {\em entire SU(2)$_L$ doublets}. Furthermore, the real scalar Higgs modes $h^0$ and $H^0$ contained in $h$ and $H$ will not be mass eigenstates, in general. On the other hand, the three Goldstone degrees of freedom $G^0, G^\pm$ for electroweak symmetry breaking are entirely contained in $h$, and only the real scalar mode $h^0$ in $h$ has couplings to $W^\pm$ and $Z$ bosons of the form $h^0 V_\mu V^\mu$. Given LHC data, we know that the VEV eigenstates are approximately the same as the mass eigenstates; in other words, we are in the ``alignment limit'' $\cos(\beta - \alpha) \ll 1$, because the light Higgs is observed to couple to particles proportional to their masses as in the SM~\cite{Craig:2013hca}. As a result we can think of the heavy Higgs boson as living mostly in $H$. The decays $H^0 \to h^0h^0$ and $H^0 \to Z_L Z_L$, where we use the Goldstone equivalence theorem to relate the decay rate to longitudinal $Z$ bosons to decays to the Goldstone mode $G^0$ inside $h$, both arise from a quartic term in the potential containing one copy of $H$ and three of $h$:
\beq
V \supset {\tilde \lambda}_1 \left(H^\dagger h + h^\dagger H\right)h^\dagger h \supset {\tilde \lambda}_1 \left(v H^0 G^+ G^- + \frac{v}{2} H^0 G^0 G^0 + \frac{3v}{2} H^0 h^0 h^0 + H^0 h^0 G^+ G^- + \ldots \right).
\label{eq:heavyhiggsdecays}
\eeq
Thus, there is a relative factor of 3 in the Feynman rule for $H^0$ to two Higgs bosons relative to $H^0$ to two Goldstones. In the first case we have three $h$ factors in the potential, one of which must be replaced by a VEV and two with a physical Higgs boson. The combinatoric factor of 3 comes from the fact that we can replace any of the three $h$'s with a VEV. In the second case we again replace one with a VEV, but the other two with Goldstones. The difference is that $H^0$ lives in the real part of $H$ and so must be paired with either an $h$ or a $v$ in the hermitian $H^\dagger h + h^\dagger H$ factor; the two Goldstones must go in the $h^\dagger h$ factor, and so we have no combinatoric freedom in this case. (Let us also mention in passing that eq.~\eqref{eq:heavyhiggsdecays} leads to several three-body decays of the heavy Higgs, suppressed by phase space but not by couplings; the phenomenology of such decays could be interesting, and is as far as we know unexplored.)

\begin{figure}[!h]
\begin{center}
\includegraphics[width=0.8 \textwidth]{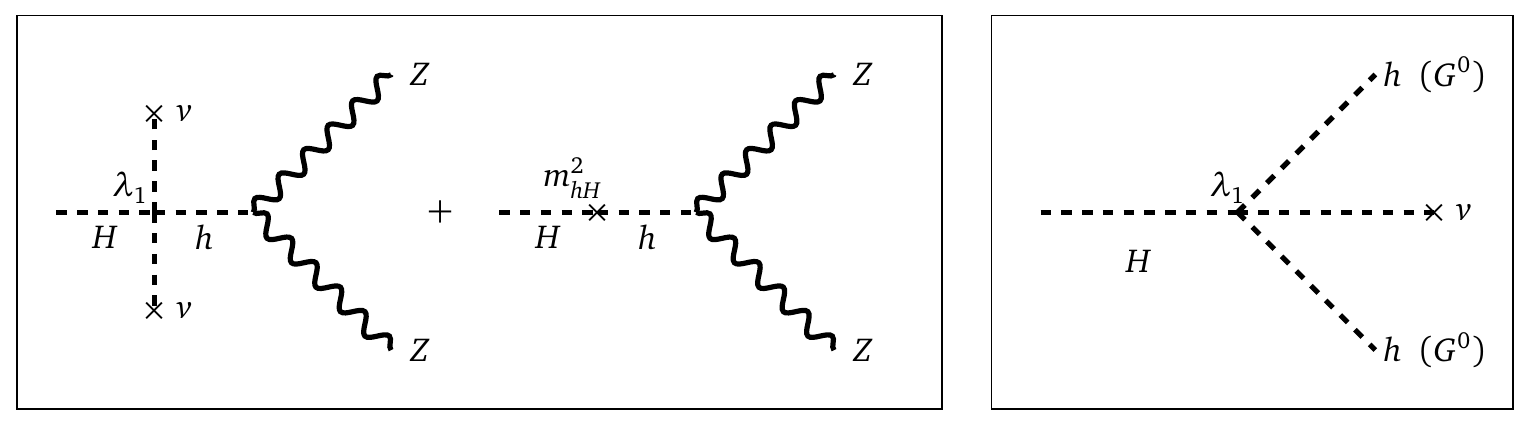}
\end{center}
\caption{Left: the decay $H \to ZZ$ in unitary gauge, for which the VEVless eigenstate $H$ first mixes into the eigenstate $h$ and then couples through its VEV to $Z_\mu Z^\mu$. Right: the decay $H \to hh$ and the related decay to two Goldstone modes. A relative factor of 3 arises from the combinatoric choice of which $h$ leg to replace by a vev in $H \to hh$. 
} \label{fig:Hdecay}
\end{figure}

The relative decay rate is also easy to understand in unitary gauge, as shown in Fig.~\ref{fig:Hdecay}. In this case another contribution arises from the mass mixing of $H$ and $h$, but this is related to the coupling ${\tilde \lambda}_1$ by a tadpole cancelation condition. In other words, our choice of $H$ as the eigenstate with zero VEV relates the terms $m_{Hh}^2 (H^\dagger h + h^\dagger H)$ and ${\tilde \lambda}_1 (H^\dagger h + h^\dagger H)(h^\dagger h)$ in the potential. In particular, the coupling for $H \to hh$ vanishes in the limit $m_{Hh}^2 \to 0$, which is the exact alignment limit where VEV eigenstates {\em are} mass eigenstates; this is reflected in the factors of $\cos(\beta - \alpha)$ in the $g_{Hhh}$ coupling in e.g. ref.~\cite{Craig:2013hca}. A little algebra shows that the unitary gauge calculation matches the Goldstone equivalence estimate up to terms of order $m_{Z,h}^2/m_H^2$, as expected on general grounds.

The case of a singlet scalar decaying to $hh$ and $ZZ$ is similar, but the combinatoric factor of 3 no longer exists, so we expect the branching ratios to be approximately equal.

{\footnotesize
\bibliography{ref}

\providecommand{\href}[2]{#2}\begingroup\raggedright\begin{thebibliography}{10%
0}

\bibitem{Barbieri:1987fn}
R.~Barbieri and G.~Giudice, ``{Upper Bounds on Supersymmetric Particle
  Masses},''
\href{http://dx.doi.org/10.1016/0550-3213(88)90171-X}{{\em Nucl.Phys.}
  {\bfseries B306} (1988) 63}.

\bibitem{Dimopoulos:1995mi}
S.~Dimopoulos and G.~Giudice, ``{Naturalness constraints in supersymmetric
  theories with nonuniversal soft terms},''
  \href{http://dx.doi.org/10.1016/0370-2693(95)00961-J}{{\em Phys.Lett.}
  {\bfseries B357} (1995) 573--578},
\href{http://arxiv.org/abs/hep-ph/9507282}{{\ttfamily arXiv:hep-ph/9507282
  [hep-ph]}}.

\bibitem{Pomarol:1995xc}
A.~Pomarol and D.~Tommasini, ``{Horizontal symmetries for the supersymmetric
  flavor problem},'' \href{http://dx.doi.org/10.1016/0550-3213(96)00074-0}{{\em
  Nucl.Phys.} {\bfseries B466} (1996) 3--24},
\href{http://arxiv.org/abs/hep-ph/9507462}{{\ttfamily arXiv:hep-ph/9507462
  [hep-ph]}}.

\bibitem{Cohen:1996vb}
A.~G. Cohen, D.~Kaplan, and A.~Nelson, ``{The More minimal supersymmetric
  standard model},''
  \href{http://dx.doi.org/10.1016/S0370-2693(96)01183-5}{{\em Phys.Lett.}
  {\bfseries B388} (1996) 588--598},
\href{http://arxiv.org/abs/hep-ph/9607394}{{\ttfamily arXiv:hep-ph/9607394
  [hep-ph]}}.

\bibitem{Kitano:2006gv}
R.~Kitano and Y.~Nomura, ``{Supersymmetry, naturalness, and signatures at the
  LHC},'' \href{http://dx.doi.org/10.1103/PhysRevD.73.095004}{{\em Phys.Rev.}
  {\bfseries D73} (2006) 095004},
\href{http://arxiv.org/abs/hep-ph/0602096}{{\ttfamily arXiv:hep-ph/0602096
  [hep-ph]}}.

\bibitem{Perelstein:2007nx}
M.~Perelstein and C.~Spethmann, ``{A Collider signature of the supersymmetric
  golden region},'' \href{http://dx.doi.org/10.1088/1126-6708/2007/04/070}{{\em
  JHEP} {\bfseries 0704} (2007) 070},
\href{http://arxiv.org/abs/hep-ph/0702038}{{\ttfamily arXiv:hep-ph/0702038
  [hep-ph]}}.

\bibitem{Papucci:2011wy}
M.~Papucci, J.~T. Ruderman, and A.~Weiler, ``{Natural SUSY Endures},''
  \href{http://dx.doi.org/10.1007/JHEP09(2012)035}{{\em JHEP} {\bfseries 1209}
  (2012) 035},
\href{http://arxiv.org/abs/1110.6926}{{\ttfamily arXiv:1110.6926 [hep-ph]}}.

\bibitem{Brust:2011tb}
C.~Brust, A.~Katz, S.~Lawrence, and R.~Sundrum, ``{SUSY, the Third Generation
  and the LHC},'' \href{http://dx.doi.org/10.1007/JHEP03(2012)103}{{\em JHEP}
  {\bfseries 1203} (2012) 103},
\href{http://arxiv.org/abs/1110.6670}{{\ttfamily arXiv:1110.6670 [hep-ph]}}.

\bibitem{Baer:2013gva}
H.~Baer, V.~Barger, and D.~Mickelson, ``{How conventional measures overestimate
  electroweak fine-tuning in supersymmetric theory},''
  \href{http://dx.doi.org/10.1103/PhysRevD.88.095013}{{\em Phys.Rev.}
  {\bfseries D88} (2013) 095013},
\href{http://arxiv.org/abs/1309.2984}{{\ttfamily arXiv:1309.2984 [hep-ph]}}.

\bibitem{Meade:2006dw}
P.~Meade and M.~Reece, ``{Top partners at the LHC: Spin and mass
  measurement},'' \href{http://dx.doi.org/10.1103/PhysRevD.74.015010}{{\em
  Phys.Rev.} {\bfseries D74} (2006) 015010},
\href{http://arxiv.org/abs/hep-ph/0601124}{{\ttfamily arXiv:hep-ph/0601124
  [hep-ph]}}.

\bibitem{Farina:2013ssa}
M.~Farina, M.~Perelstein, and N.~R.-L. Lorier, ``{Higgs Couplings and
  Naturalness},''
\href{http://arxiv.org/abs/1305.6068}{{\ttfamily arXiv:1305.6068 [hep-ph]}}.

\bibitem{Kribs:2013lua}
G.~D. Kribs, A.~Martin, and A.~Menon, ``{Natural Supersymmetry and Implications
  for Higgs physics},''
  \href{http://dx.doi.org/10.1103/PhysRevD.88.035025}{{\em Phys.Rev.}
  {\bfseries D88} (2013) 035025},
\href{http://arxiv.org/abs/1305.1313}{{\ttfamily arXiv:1305.1313 [hep-ph]}}.

\bibitem{Kowalska:2013ica}
K.~Kowalska and E.~M. Sessolo, ``{Natural MSSM after the LHC 8 TeV run},''
  \href{http://dx.doi.org/10.1103/PhysRevD.88.075001}{{\em Phys.Rev.}
  {\bfseries D88} (2013) 075001},
\href{http://arxiv.org/abs/1307.5790}{{\ttfamily arXiv:1307.5790 [hep-ph]}}.

\bibitem{Han:2013kga}
C.~Han, K.-I. Hikasa, L.~Wu, J.~M. Yang, and Y.~Zhang, ``{Current experimental
  bounds on stop mass in natural SUSY},''
  \href{http://dx.doi.org/10.1007/JHEP10(2013)216}{{\em JHEP} {\bfseries 1310}
  (2013) 216},
\href{http://arxiv.org/abs/1308.5307}{{\ttfamily arXiv:1308.5307 [hep-ph]}}.

\bibitem{Craig:2013cxa}
N.~Craig, ``{The State of Supersymmetry after Run I of the LHC},''
\href{http://arxiv.org/abs/1309.0528}{{\ttfamily arXiv:1309.0528 [hep-ph]}}.

\bibitem{Feng:2013pwa}
J.~L. Feng, ``{Naturalness and the Status of Supersymmetry},''
  \href{http://dx.doi.org/10.1146/annurev-nucl-102010-130447}{{\em
  Ann.Rev.Nucl.Part.Sci.} {\bfseries 63} (2013) 351--382},
\href{http://arxiv.org/abs/1302.6587}{{\ttfamily arXiv:1302.6587 [hep-ph]}}.

\bibitem{Hardy:2013ywa}
E.~Hardy, ``{Is Natural SUSY Natural?},''
  \href{http://dx.doi.org/10.1007/JHEP10(2013)133}{{\em JHEP} {\bfseries 1310}
  (2013) 133},
\href{http://arxiv.org/abs/1306.1534}{{\ttfamily arXiv:1306.1534 [hep-ph]}}.

\bibitem{Batra:2003nj}
P.~Batra, A.~Delgado, D.~E. Kaplan, and T.~M. Tait, ``{The Higgs mass bound in
  gauge extensions of the minimal supersymmetric standard model},''
  \href{http://dx.doi.org/10.1088/1126-6708/2004/02/043}{{\em JHEP} {\bfseries
  0402} (2004) 043},
\href{http://arxiv.org/abs/hep-ph/0309149}{{\ttfamily arXiv:hep-ph/0309149
  [hep-ph]}}.

\bibitem{Barbieri:2006bg}
R.~Barbieri, L.~J. Hall, Y.~Nomura, and V.~S. Rychkov, ``{Supersymmetry without
  a Light Higgs Boson},''
  \href{http://dx.doi.org/10.1103/PhysRevD.75.035007}{{\em Phys.Rev.}
  {\bfseries D75} (2007) 035007},
\href{http://arxiv.org/abs/hep-ph/0607332}{{\ttfamily arXiv:hep-ph/0607332
  [hep-ph]}}.

\bibitem{Dine:2007xi}
M.~Dine, N.~Seiberg, and S.~Thomas, ``{Higgs physics as a window beyond the
  MSSM (BMSSM)},'' \href{http://dx.doi.org/10.1103/PhysRevD.76.095004}{{\em
  Phys.Rev.} {\bfseries D76} (2007) 095004},
\href{http://arxiv.org/abs/0707.0005}{{\ttfamily arXiv:0707.0005 [hep-ph]}}.

\bibitem{Hall:2011aa}
L.~J. Hall, D.~Pinner, and J.~T. Ruderman, ``{A Natural SUSY Higgs Near 126
  GeV},'' \href{http://dx.doi.org/10.1007/JHEP04(2012)131}{{\em JHEP}
  {\bfseries 1204} (2012) 131},
\href{http://arxiv.org/abs/1112.2703}{{\ttfamily arXiv:1112.2703 [hep-ph]}}.

\bibitem{Gherghetta:2014xea}
T.~Gherghetta, B.~von Harling, A.~D. Medina, and M.~A. Schmidt, ``{The price of
  being SM-like in SUSY},''
  \href{http://dx.doi.org/10.1007/JHEP04(2014)180}{{\em JHEP} {\bfseries 1404}
  (2014) 180},
\href{http://arxiv.org/abs/1401.8291}{{\ttfamily arXiv:1401.8291 [hep-ph]}}.

\bibitem{Ishiwata:2011ab}
K.~Ishiwata, N.~Nagata, and N.~Yokozaki, ``{Natural Supersymmetry and $b \to s
  \gamma$ constraints},''
  \href{http://dx.doi.org/10.1016/j.physletb.2012.02.052}{{\em Phys.Lett.}
  {\bfseries B710} (2012) 145--148},
\href{http://arxiv.org/abs/1112.1944}{{\ttfamily arXiv:1112.1944 [hep-ph]}}.

\bibitem{Gunion:2002zf}
J.~F. Gunion and H.~E. Haber, ``{The CP conserving two Higgs doublet model: The
  Approach to the decoupling limit},''
  \href{http://dx.doi.org/10.1103/PhysRevD.67.075019}{{\em Phys.Rev.}
  {\bfseries D67} (2003) 075019},
\href{http://arxiv.org/abs/hep-ph/0207010}{{\ttfamily arXiv:hep-ph/0207010
  [hep-ph]}}.

\bibitem{Randall:2007as}
L.~Randall, ``{Two Higgs Models for Large Tan Beta and Heavy Second Higgs},''
  \href{http://dx.doi.org/10.1088/1126-6708/2008/02/084}{{\em JHEP} {\bfseries
  0802} (2008) 084},
\href{http://arxiv.org/abs/0711.4360}{{\ttfamily arXiv:0711.4360 [hep-ph]}}.

\bibitem{Blum:2012kn}
K.~Blum and R.~T. D'Agnolo, ``{2 Higgs or not 2 Higgs},''
  \href{http://dx.doi.org/10.1016/j.physletb.2012.06.054}{{\em Phys.Lett.}
  {\bfseries B714} (2012) 66--69},
\href{http://arxiv.org/abs/1202.2364}{{\ttfamily arXiv:1202.2364 [hep-ph]}}.

\bibitem{Lu:2013cta}
X.~Lu, H.~Murayama, J.~T. Ruderman, and K.~Tobioka, ``{A Natural Higgs Mass in
  Supersymmetry from Non-Decoupling Effects},''
  \href{http://dx.doi.org/10.1103/PhysRevLett.112.191803}{{\em Phys.Rev.Lett.}
  {\bfseries 112} (2014) 191803},
\href{http://arxiv.org/abs/1308.0792}{{\ttfamily arXiv:1308.0792 [hep-ph]}}.

\bibitem{Maloney:2004rc}
A.~Maloney, A.~Pierce, and J.~G. Wacker, ``{D-terms, unification, and the Higgs
  mass},'' \href{http://dx.doi.org/10.1088/1126-6708/2006/06/034}{{\em JHEP}
  {\bfseries 0606} (2006) 034},
\href{http://arxiv.org/abs/hep-ph/0409127}{{\ttfamily arXiv:hep-ph/0409127
  [hep-ph]}}.

\bibitem{Cheung:2012zq}
C.~Cheung and H.~L. Roberts, ``{Higgs Mass from D-Terms: a Litmus Test},''
  \href{http://dx.doi.org/10.1007/JHEP12(2013)018}{{\em JHEP} {\bfseries 1312}
  (2013) 018},
\href{http://arxiv.org/abs/1207.0234}{{\ttfamily arXiv:1207.0234 [hep-ph]}}.

\bibitem{Espinosa:1991gr}
J.~Espinosa and M.~Quiros, ``{On Higgs boson masses in nonminimal
  supersymmetric standard models},''
\href{http://dx.doi.org/10.1016/0370-2693(92)91846-2}{{\em Phys.Lett.}
  {\bfseries B279} (1992) 92--97}.

\bibitem{Agashe:2012zq}
K.~Agashe, Y.~Cui, and R.~Franceschini, ``{Natural Islands for a 125 GeV Higgs
  in the scale-invariant NMSSM},''
  \href{http://dx.doi.org/10.1007/JHEP02(2013)031}{{\em JHEP} {\bfseries 1302}
  (2013) 031},
\href{http://arxiv.org/abs/1209.2115}{{\ttfamily arXiv:1209.2115 [hep-ph]}}.

\bibitem{Gherghetta:2012gb}
T.~Gherghetta, B.~von Harling, A.~D. Medina, and M.~A. Schmidt, ``{The
  Scale-Invariant NMSSM and the 126 GeV Higgs Boson},''
  \href{http://dx.doi.org/10.1007/JHEP02(2013)032}{{\em JHEP} {\bfseries 02}
  (2013) 032},
\href{http://arxiv.org/abs/1212.5243}{{\ttfamily arXiv:1212.5243 [hep-ph]}}.

\bibitem{Agashe:2011ia}
K.~Agashe, A.~Azatov, A.~Katz, and D.~Kim, ``{Improving the tunings of the MSSM
  by adding triplets and singlet},''
  \href{http://dx.doi.org/10.1103/PhysRevD.84.115024}{{\em Phys.Rev.}
  {\bfseries D84} (2011) 115024},
\href{http://arxiv.org/abs/1109.2842}{{\ttfamily arXiv:1109.2842 [hep-ph]}}.

\bibitem{Baer:2014ica}
H.~Baer, V.~Barger, D.~Mickelson, and M.~Padeffke-Kirkland, ``{SUSY models
  under siege: LHC constraints and electroweak fine-tuning},''
\href{http://arxiv.org/abs/1404.2277}{{\ttfamily arXiv:1404.2277 [hep-ph]}}.

\bibitem{Fichet:2012sn}
S.~Fichet, ``{Quantified naturalness from Bayesian statistics},''
  \href{http://dx.doi.org/10.1103/PhysRevD.86.125029}{{\em Phys.Rev.}
  {\bfseries D86} (2012) 125029},
\href{http://arxiv.org/abs/1204.4940}{{\ttfamily arXiv:1204.4940 [hep-ph]}}.

\bibitem{Fan:2014txa}
J.~Fan and M.~Reece, ``{A New Look at Higgs Constraints on Stops},''
\href{http://arxiv.org/abs/1401.7671}{{\ttfamily arXiv:1401.7671 [hep-ph]}}.

\bibitem{Haber:1990aw}
H.~E. Haber and R.~Hempfling, ``{Can the mass of the lightest Higgs boson of
  the minimal supersymmetric model be larger than m(Z)?},''
\href{http://dx.doi.org/10.1103/PhysRevLett.66.1815}{{\em Phys.Rev.Lett.}
  {\bfseries 66} (1991) 1815--1818}.

\bibitem{Barbieri:1990ja}
R.~Barbieri, M.~Frigeni, and F.~Caravaglios, ``{The Supersymmetric Higgs for
  heavy superpartners},''
\href{http://dx.doi.org/10.1016/0370-2693(91)91226-L}{{\em Phys.Lett.}
  {\bfseries B258} (1991) 167--170}.

\bibitem{Casas:1994us}
J.~Casas, J.~Espinosa, M.~Quiros, and A.~Riotto, ``{The Lightest Higgs boson
  mass in the minimal supersymmetric standard model},''
  \href{http://dx.doi.org/10.1016/0550-3213(94)00508-C}{{\em Nucl.Phys.}
  {\bfseries B436} (1995) 3--29},
\href{http://arxiv.org/abs/hep-ph/9407389}{{\ttfamily arXiv:hep-ph/9407389
  [hep-ph]}}.

\bibitem{Carena:1995bx}
M.~S. Carena, J.~Espinosa, M.~Quiros, and C.~Wagner, ``{Analytical expressions
  for radiatively corrected Higgs masses and couplings in the MSSM},''
  \href{http://dx.doi.org/10.1016/0370-2693(95)00694-G}{{\em Phys.Lett.}
  {\bfseries B355} (1995) 209--221},
\href{http://arxiv.org/abs/hep-ph/9504316}{{\ttfamily arXiv:hep-ph/9504316
  [hep-ph]}}.

\bibitem{Craig:2011yk}
N.~Craig, D.~Green, and A.~Katz, ``{(De)Constructing a Natural and Flavorful
  Supersymmetric Standard Model},''
  \href{http://dx.doi.org/10.1007/JHEP07(2011)045}{{\em JHEP} {\bfseries 1107}
  (2011) 045},
\href{http://arxiv.org/abs/1103.3708}{{\ttfamily arXiv:1103.3708 [hep-ph]}}.

\bibitem{Craig:2012bs}
N.~Craig and A.~Katz, ``{A Supersymmetric Higgs Sector with Chiral D-terms},''
  \href{http://dx.doi.org/10.1007/JHEP05(2013)015}{{\em JHEP} {\bfseries 1305}
  (2013) 015},
\href{http://arxiv.org/abs/1212.2635}{{\ttfamily arXiv:1212.2635 [hep-ph]}}.

\bibitem{Bharucha:2013ela}
A.~Bharucha, A.~Goudelis, and M.~McGarrie, ``{En-gauging Naturalness},''
\href{http://arxiv.org/abs/1310.4500}{{\ttfamily arXiv:1310.4500 [hep-ph]}}.

\bibitem{CMS-PAS-HIG-13-021}
{\bfseries CMS} Collaboration, ``{Higgs to tau tau (MSSM)},'' Tech. Rep.
  CMS-PAS-HIG-13-021, CERN, Geneva, 2013.
\newblock \url{http://cds.cern.ch/record/1623367}.

\bibitem{Gupta:2012fy}
R.~S. Gupta, M.~Montull, and F.~Riva, ``{SUSY Faces its Higgs Couplings},''
  \href{http://dx.doi.org/10.1007/JHEP04(2013)132}{{\em JHEP} {\bfseries 1304}
  (2013) 132},
\href{http://arxiv.org/abs/1212.5240}{{\ttfamily arXiv:1212.5240 [hep-ph]}}.

\bibitem{Blum:2012ii}
K.~Blum, R.~T. D'Agnolo, and J.~Fan, ``{Natural SUSY Predicts: Higgs
  Couplings},'' \href{http://dx.doi.org/10.1007/JHEP01(2013)057}{{\em JHEP}
  {\bfseries 1301} (2013) 057},
\href{http://arxiv.org/abs/1206.5303}{{\ttfamily arXiv:1206.5303 [hep-ph]}}.

\bibitem{Espinosa:2012in}
J.~R. Espinosa, C.~Grojean, V.~Sanz, and M.~Trott, ``{NSUSY fits},''
  \href{http://dx.doi.org/10.1007/JHEP12(2012)077}{{\em JHEP} {\bfseries 1212}
  (2012) 077},
\href{http://arxiv.org/abs/1207.7355}{{\ttfamily arXiv:1207.7355 [hep-ph]}}.

\bibitem{Oshimo:1992zd}
N.~Oshimo, ``{Radiative $B$ meson decay in supersymmetric models},''
\href{http://dx.doi.org/10.1016/0550-3213(93)90471-Z}{{\em Nucl.Phys.}
  {\bfseries B404} (1993) 20--41}.

\bibitem{Barbieri:1993av}
R.~Barbieri and G.~Giudice, ``{$b\to s \gamma$ decay and supersymmetry},''
  \href{http://dx.doi.org/10.1016/0370-2693(93)91508-K}{{\em Phys.Lett.}
  {\bfseries B309} (1993) 86--90},
\href{http://arxiv.org/abs/hep-ph/9303270}{{\ttfamily arXiv:hep-ph/9303270
  [hep-ph]}}.

\bibitem{Okada:1993sx}
Y.~Okada, ``{Light stop and the $b \to s \gamma$ process},''
  \href{http://dx.doi.org/10.1016/0370-2693(93)90167-G}{{\em Phys.Lett.}
  {\bfseries B315} (1993) 119--123},
\href{http://arxiv.org/abs/hep-ph/9307249}{{\ttfamily arXiv:hep-ph/9307249
  [hep-ph]}}.

\bibitem{Garisto:1993jc}
R.~Garisto and J.~Ng, ``{Supersymmetric $b \to s \gamma$ with large chargino
  contributions},'' \href{http://dx.doi.org/10.1016/0370-2693(93)91627-Y}{{\em
  Phys.Lett.} {\bfseries B315} (1993) 372--378},
\href{http://arxiv.org/abs/hep-ph/9307301}{{\ttfamily arXiv:hep-ph/9307301
  [hep-ph]}}.

\bibitem{Baer:1997jq}
H.~Baer, M.~Brhlik, D.~Castano, and X.~Tata, ``{$b \to s \gamma$ constraints on
  the minimal supergravity model with large tan Beta},''
  \href{http://dx.doi.org/10.1103/PhysRevD.58.015007}{{\em Phys.Rev.}
  {\bfseries D58} (1998) 015007},
\href{http://arxiv.org/abs/hep-ph/9712305}{{\ttfamily arXiv:hep-ph/9712305
  [hep-ph]}}.

\bibitem{Degrassi:2000qf}
G.~Degrassi, P.~Gambino, and G.~Giudice, ``{$B \to X_s \gamma$ in
  supersymmetry: Large contributions beyond the leading order},''
  \href{http://dx.doi.org/10.1088/1126-6708/2000/12/009}{{\em JHEP} {\bfseries
  0012} (2000) 009},
\href{http://arxiv.org/abs/hep-ph/0009337}{{\ttfamily arXiv:hep-ph/0009337
  [hep-ph]}}.

\bibitem{Carena:2000uj}
M.~S. Carena, D.~Garcia, U.~Nierste, and C.~E. Wagner, ``{$b\to s \gamma$ and
  supersymmetry with large tan $\beta$},''
  \href{http://dx.doi.org/10.1016/S0370-2693(01)00009-0}{{\em Phys.Lett.}
  {\bfseries B499} (2001) 141--146},
\href{http://arxiv.org/abs/hep-ph/0010003}{{\ttfamily arXiv:hep-ph/0010003
  [hep-ph]}}.

\bibitem{Heister:2002mn}
{\bfseries ALEPH} Collaboration, A.~Heister {\em et~al.}, ``{Search for
  charginos nearly mass degenerate with the lightest neutralino in e+ e-
  collisions at center-of-mass energies up to 209-GeV},''
  \href{http://dx.doi.org/10.1016/S0370-2693(02)01584-8}{{\em Phys.Lett.}
  {\bfseries B533} (2002) 223--236},
\href{http://arxiv.org/abs/hep-ex/0203020}{{\ttfamily arXiv:hep-ex/0203020
  [hep-ex]}}.

\bibitem{Abdallah:2003xe}
{\bfseries DELPHI} Collaboration, J.~Abdallah {\em et~al.}, ``{Searches for
  supersymmetric particles in $e^+ e^-$ collisions up to 208-GeV and
  interpretation of the results within the MSSM},''
  \href{http://dx.doi.org/10.1140/epjc/s2003-01355-5}{{\em Eur.Phys.J.}
  {\bfseries C31} (2003) 421--479},
\href{http://arxiv.org/abs/hep-ex/0311019}{{\ttfamily arXiv:hep-ex/0311019
  [hep-ex]}}.

\bibitem{Abbiendi:2002vz}
{\bfseries OPAL} Collaboration, G.~Abbiendi {\em et~al.}, ``{Search for nearly
  mass degenerate charginos and neutralinos at LEP},''
  \href{http://dx.doi.org/10.1140/epjc/s2003-01237-x}{{\em Eur.Phys.J.}
  {\bfseries C29} (2003) 479--489},
\href{http://arxiv.org/abs/hep-ex/0210043}{{\ttfamily arXiv:hep-ex/0210043
  [hep-ex]}}.

\bibitem{Acciarri:2000wy}
{\bfseries L3} Collaboration, M.~Acciarri {\em et~al.}, ``{Search for charginos
  with a small mass difference with the lightest supersymmetric particle at
  $\sqrt{s}$ = 189~GeV},''
  \href{http://dx.doi.org/10.1016/S0370-2693(00)00488-3}{{\em Phys.Lett.}
  {\bfseries B482} (2000) 31--42},
\href{http://arxiv.org/abs/hep-ex/0002043}{{\ttfamily arXiv:hep-ex/0002043
  [hep-ex]}}.

\bibitem{Gori:2013ala}
S.~Gori, S.~Jung, and L.-T. Wang, ``{Cornering electroweakinos at the LHC},''
  \href{http://dx.doi.org/10.1007/JHEP10(2013)191}{{\em JHEP} {\bfseries 1310}
  (2013) 191},
\href{http://arxiv.org/abs/1307.5952}{{\ttfamily arXiv:1307.5952 [hep-ph]}}.

\bibitem{Han:2013usa}
C.~Han, A.~Kobakhidze, N.~Liu, A.~Saavedra, L.~Wu, and J.~M. Yang, ``{Probing
  Light Higgsinos in Natural SUSY from Monojet Signals at the LHC},''
  \href{http://dx.doi.org/10.1007/JHEP02(2014)049}{{\em JHEP} {\bfseries 1402}
  (2014) 049},
\href{http://arxiv.org/abs/1310.4274}{{\ttfamily arXiv:1310.4274 [hep-ph]}}.

\bibitem{Schwaller:2013baa}
P.~Schwaller and J.~Zurita, ``{Compressed electroweakino spectra at the LHC},''
  \href{http://dx.doi.org/10.1007/JHEP03(2014)060}{{\em JHEP} {\bfseries 1403}
  (2014) 060},
\href{http://arxiv.org/abs/1312.7350}{{\ttfamily arXiv:1312.7350 [hep-ph]}}.

\bibitem{Han:2014kaa}
Z.~Han, G.~D. Kribs, A.~Martin, and A.~Menon, ``{Hunting Quasi-Degenerate
  Higgsinos},'' \href{http://dx.doi.org/10.1103/PhysRevD.89.075007}{{\em
  Phys.Rev.} {\bfseries D89} (2014) 075007},
\href{http://arxiv.org/abs/1401.1235}{{\ttfamily arXiv:1401.1235 [hep-ph]}}.

\bibitem{Aad:2013wta}
{\bfseries ATLAS} Collaboration, G.~Aad {\em et~al.}, ``{Search for new
  phenomena in final states with large jet multiplicities and missing
  transverse momentum at $\sqrt{s}=8$~TeV proton-proton collisions using the
  ATLAS experiment},'' \href{http://dx.doi.org/10.1007/JHEP10(2013)130}{{\em
  JHEP} {\bfseries 1310} (2013) 130},
\href{http://arxiv.org/abs/1308.1841}{{\ttfamily arXiv:1308.1841 [hep-ex]}}.

\bibitem{Chatrchyan:2013iqa}
{\bfseries CMS} Collaboration, S.~Chatrchyan {\em et~al.}, ``{Search for
  supersymmetry in pp collisions at $\sqrt{s}$ = 8 TeV in events with a single
  lepton, large jet multiplicity, and multiple b jets},''
\href{http://arxiv.org/abs/1311.4937}{{\ttfamily arXiv:1311.4937 [hep-ex]}}.

\bibitem{Chatrchyan:2014lfa}
{\bfseries CMS} Collaboration, S.~Chatrchyan {\em et~al.}, ``{Search for new
  physics in the multijet and missing transverse momentum final state in
  proton-proton collisions at $\sqrt{s}$ = 8 TeV},''
\href{http://arxiv.org/abs/1402.4770}{{\ttfamily arXiv:1402.4770 [hep-ex]}}.

\bibitem{TheATLAScollaboration:2013xia}
{\bfseries ATLAS} Collaboration, ``{Search for massive particles in multijet
  signatures with the ATLAS detector in $\sqrt{s} = 8$ TeV pp collisions at the
  LHC},''.
\url{http://cds.cern.ch/record/1595753}.

\bibitem{Evans:2013jna}
J.~A. Evans, Y.~Kats, D.~Shih, and M.~J. Strassler, ``{Toward Full LHC Coverage
  of Natural Supersymmetry},''
\href{http://arxiv.org/abs/1310.5758}{{\ttfamily arXiv:1310.5758 [hep-ph]}}.

\bibitem{Altmannshofer:2012ks}
W.~Altmannshofer, M.~Carena, N.~R. Shah, and F.~Yu, ``{Indirect Probes of the
  MSSM after the Higgs Discovery},''
  \href{http://dx.doi.org/10.1007/JHEP01(2013)160}{{\em JHEP} {\bfseries 1301}
  (2013) 160},
\href{http://arxiv.org/abs/1211.1976}{{\ttfamily arXiv:1211.1976 [hep-ph]}}.

\bibitem{Freitas:2008vh}
A.~Freitas and U.~Haisch, ``{$\overline{B} \to X_s \gamma$ gamma in two
  universal extra dimensions},''
  \href{http://dx.doi.org/10.1103/PhysRevD.77.093008}{{\em Phys.Rev.}
  {\bfseries D77} (2008) 093008},
\href{http://arxiv.org/abs/0801.4346}{{\ttfamily arXiv:0801.4346 [hep-ph]}}.

\bibitem{Mescia:2012fg}
F.~Mescia and J.~Virto, ``{Natural SUSY and Kaon Mixing in view of recent
  results from Lattice QCD},''
  \href{http://dx.doi.org/10.1103/PhysRevD.86.095004}{{\em Phys.Rev.}
  {\bfseries D86} (2012) 095004},
\href{http://arxiv.org/abs/1208.0534}{{\ttfamily arXiv:1208.0534 [hep-ph]}}.

\bibitem{Martin:1997ns}
S.~P. Martin, ``{A Supersymmetry primer},''
\href{http://arxiv.org/abs/hep-ph/9709356}{{\ttfamily arXiv:hep-ph/9709356
  [hep-ph]}}.

\bibitem{Dobrescu:2010mk}
B.~A. Dobrescu and P.~J. Fox, ``{Uplifted supersymmetric Higgs region},''
  \href{http://dx.doi.org/10.1140/epjc/s10052-010-1399-y}{{\em Eur.Phys.J.}
  {\bfseries C70} (2010) 263--270},
\href{http://arxiv.org/abs/1001.3147}{{\ttfamily arXiv:1001.3147 [hep-ph]}}.

\bibitem{Altmannshofer:2010zt}
W.~Altmannshofer and D.~M. Straub, ``{Viability of MSSM scenarios at very large
  tan(beta)},'' \href{http://dx.doi.org/10.1007/JHEP09(2010)078}{{\em JHEP}
  {\bfseries 1009} (2010) 078},
\href{http://arxiv.org/abs/1004.1993}{{\ttfamily arXiv:1004.1993 [hep-ph]}}.

\bibitem{Kusenko:1996jn}
A.~Kusenko, P.~Langacker, and G.~Segre, ``{Phase transitions and vacuum
  tunneling into charge and color breaking minima in the MSSM},''
  \href{http://dx.doi.org/10.1103/PhysRevD.54.5824}{{\em Phys.Rev.} {\bfseries
  D54} (1996) 5824--5834},
\href{http://arxiv.org/abs/hep-ph/9602414}{{\ttfamily arXiv:hep-ph/9602414
  [hep-ph]}}.

\bibitem{Misiak:2006zs}
M.~Misiak, H.~Asatrian, K.~Bieri, M.~Czakon, A.~Czarnecki, {\em et~al.},
  ``{Estimate of Br$(B \to X_s \gamma)$ at O($\alpha_s^2$)},''
  \href{http://dx.doi.org/10.1103/PhysRevLett.98.022002}{{\em Phys.Rev.Lett.}
  {\bfseries 98} (2007) 022002},
\href{http://arxiv.org/abs/hep-ph/0609232}{{\ttfamily arXiv:hep-ph/0609232
  [hep-ph]}}.

\bibitem{Lees:2012ufa}
{\bfseries BaBar} Collaboration, J.~Lees {\em et~al.}, ``{Measurement of
  B($B\to X_s \gamma$), the $B\to X_s \gamma$ photon energy spectrum, and the
  direct CP asymmetry in $B\to X_{s+d} \gamma$ decays},''
  \href{http://dx.doi.org/10.1103/PhysRevD.86.112008}{{\em Phys.Rev.}
  {\bfseries D86} (2012) 112008},
\href{http://arxiv.org/abs/1207.5772}{{\ttfamily arXiv:1207.5772 [hep-ex]}}.

\bibitem{Amhis:2012bh}
{\bfseries Heavy Flavor Averaging Group} Collaboration, Y.~Amhis {\em et~al.},
  ``{Averages of B-Hadron, C-Hadron, and tau-lepton properties as of early
  2012},''
\href{http://arxiv.org/abs/1207.1158}{{\ttfamily arXiv:1207.1158 [hep-ex]}}.

\bibitem{Babu:1999hn}
K.~Babu and C.~F. Kolda, ``{Higgs mediated $B^0 \to \mu^{+} \mu^{-}$ in minimal
  supersymmetry},'' \href{http://dx.doi.org/10.1103/PhysRevLett.84.228}{{\em
  Phys.Rev.Lett.} {\bfseries 84} (2000) 228--231},
\href{http://arxiv.org/abs/hep-ph/9909476}{{\ttfamily arXiv:hep-ph/9909476
  [hep-ph]}}.

\bibitem{Isidori:2001fv}
G.~Isidori and A.~Retico, ``{Scalar flavor changing neutral currents in the
  large tan beta limit},''
  \href{http://dx.doi.org/10.1088/1126-6708/2001/11/001}{{\em JHEP} {\bfseries
  0111} (2001) 001},
\href{http://arxiv.org/abs/hep-ph/0110121}{{\ttfamily arXiv:hep-ph/0110121
  [hep-ph]}}.

\bibitem{Aaij:2012nna}
{\bfseries LHCb} Collaboration, R.~Aaij {\em et~al.}, ``{First Evidence for the
  Decay $B^0_s \to \mu^+\mu^-$},''
  \href{http://dx.doi.org/10.1103/PhysRevLett.110.021801}{{\em Phys.Rev.Lett.}
  {\bfseries 110} (2013) 021801},
\href{http://arxiv.org/abs/1211.2674}{{\ttfamily arXiv:1211.2674 [hep-ex]}}.

\bibitem{Buras:2012ru}
A.~J. Buras, J.~Girrbach, D.~Guadagnoli, and G.~Isidori, ``{On the Standard
  Model prediction for $BR(B_{s,d} \to \mu^+ \mu^-)$},''
  \href{http://dx.doi.org/10.1140/epjc/s10052-012-2172-1}{{\em Eur.Phys.J.}
  {\bfseries C72} (2012) 2172},
\href{http://arxiv.org/abs/1208.0934}{{\ttfamily arXiv:1208.0934 [hep-ph]}}.

\bibitem{Chankowski:2000ng}
P.~H. Chankowski and L.~Slawianowska, ``{$B^0_{d,s} \to mu^- mu^+$ decay in the
  MSSM},'' \href{http://dx.doi.org/10.1103/PhysRevD.63.054012}{{\em Phys.Rev.}
  {\bfseries D63} (2001) 054012},
\href{http://arxiv.org/abs/hep-ph/0008046}{{\ttfamily arXiv:hep-ph/0008046
  [hep-ph]}}.

\bibitem{Fox:2002bu}
P.~J. Fox, A.~E. Nelson, and N.~Weiner, ``{Dirac gaugino masses and supersoft
  supersymmetry breaking},''
  \href{http://dx.doi.org/10.1088/1126-6708/2002/08/035}{{\em JHEP} {\bfseries
  0208} (2002) 035},
\href{http://arxiv.org/abs/hep-ph/0206096}{{\ttfamily arXiv:hep-ph/0206096
  [hep-ph]}}.

\bibitem{Nelson:2002ca}
A.~E. Nelson, N.~Rius, V.~Sanz, and M.~Unsal, ``{The Minimal supersymmetric
  model without a mu term},''
  \href{http://dx.doi.org/10.1088/1126-6708/2002/08/039}{{\em JHEP} {\bfseries
  0208} (2002) 039},
\href{http://arxiv.org/abs/hep-ph/0206102}{{\ttfamily arXiv:hep-ph/0206102
  [hep-ph]}}.

\bibitem{Kribs:2007ac}
G.~D. Kribs, E.~Poppitz, and N.~Weiner, ``{Flavor in supersymmetry with an
  extended R-symmetry},''
  \href{http://dx.doi.org/10.1103/PhysRevD.78.055010}{{\em Phys.Rev.}
  {\bfseries D78} (2008) 055010},
\href{http://arxiv.org/abs/0712.2039}{{\ttfamily arXiv:0712.2039 [hep-ph]}}.

\bibitem{Davies:2011mp}
R.~Davies, J.~March-Russell, and M.~McCullough, ``{A Supersymmetric One Higgs
  Doublet Model},'' \href{http://dx.doi.org/10.1007/JHEP04(2011)108}{{\em JHEP}
  {\bfseries 1104} (2011) 108},
\href{http://arxiv.org/abs/1103.1647}{{\ttfamily arXiv:1103.1647 [hep-ph]}}.

\bibitem{Frugiuele:2011mh}
C.~Frugiuele and T.~Gregoire, ``{Making the Sneutrino a Higgs with a $U(1)_R$
  Lepton Number},'' \href{http://dx.doi.org/10.1103/PhysRevD.85.015016}{{\em
  Phys.Rev.} {\bfseries D85} (2012) 015016},
\href{http://arxiv.org/abs/1107.4634}{{\ttfamily arXiv:1107.4634 [hep-ph]}}.

\bibitem{Kribs:2012gx}
G.~D. Kribs and A.~Martin, ``{Supersoft Supersymmetry is Super-Safe},''
  \href{http://dx.doi.org/10.1103/PhysRevD.85.115014}{{\em Phys.Rev.}
  {\bfseries D85} (2012) 115014},
\href{http://arxiv.org/abs/1203.4821}{{\ttfamily arXiv:1203.4821 [hep-ph]}}.

\bibitem{Benakli:2012cy}
K.~Benakli, M.~D. Goodsell, and F.~Staub, ``{Dirac Gauginos and the 125 GeV
  Higgs},'' \href{http://dx.doi.org/10.1007/JHEP06(2013)073}{{\em JHEP}
  {\bfseries 1306} (2013) 073},
\href{http://arxiv.org/abs/1211.0552}{{\ttfamily arXiv:1211.0552 [hep-ph]}}.

\bibitem{Han:2012cu}
Z.~Han, A.~Katz, M.~Son, and B.~Tweedie, ``{Boosting Searches for Natural SUSY
  with RPV via Gluino Cascades},''
  \href{http://dx.doi.org/10.1103/PhysRevD.87.075003}{{\em Phys.Rev.}
  {\bfseries D87} (2013) 075003},
\href{http://arxiv.org/abs/1211.4025}{{\ttfamily arXiv:1211.4025 [hep-ph]}}.

\bibitem{Arvanitaki:2013yja}
A.~Arvanitaki, M.~Baryakhtar, X.~Huang, K.~van Tilburg, and G.~Villadoro,
  ``{The Last Vestiges of Naturalness},''
  \href{http://dx.doi.org/10.1007/JHEP03(2014)022}{{\em JHEP} {\bfseries 1403}
  (2014) 022},
\href{http://arxiv.org/abs/1309.3568}{{\ttfamily arXiv:1309.3568 [hep-ph]}}.

\bibitem{Csaki:2013fla}
C.~Csaki, J.~Goodman, R.~Pavesi, and Y.~Shirman, ``{The $m_D-b_M$ Problem of
  Dirac Gauginos and its Solutions},''
  \href{http://dx.doi.org/10.1103/PhysRevD.89.055005}{{\em Phys.Rev.}
  {\bfseries D89} (2014) 055005},
\href{http://arxiv.org/abs/1310.4504}{{\ttfamily arXiv:1310.4504 [hep-ph]}}.

\bibitem{Kribs:2013eua}
G.~D. Kribs and N.~Raj, ``{Mixed Gauginos Sending Mixed Messages to the LHC},''
  \href{http://dx.doi.org/10.1103/PhysRevD.89.055011}{{\em Phys.Rev.}
  {\bfseries D89} (2014) 055011},
\href{http://arxiv.org/abs/1307.7197}{{\ttfamily arXiv:1307.7197 [hep-ph]}}.

\bibitem{Bertuzzo:2014bwa}
E.~Bertuzzo, C.~Frugiuele, T.~Gregoire, and E.~Ponton, ``{Dirac gauginos, R
  symmetry and the 125 GeV Higgs},''
\href{http://arxiv.org/abs/1402.5432}{{\ttfamily arXiv:1402.5432 [hep-ph]}}.

\bibitem{Arbey:2013jla}
A.~Arbey, M.~Battaglia, and F.~Mahmoudi, ``{Supersymmetric Heavy Higgs Bosons
  at the LHC},'' \href{http://dx.doi.org/10.1103/PhysRevD.88.015007}{{\em
  Phys.Rev.} {\bfseries D88} (2013) 015007},
\href{http://arxiv.org/abs/1303.7450}{{\ttfamily arXiv:1303.7450 [hep-ph]}}.

\bibitem{Brownson:2013lka}
E.~Brownson, N.~Craig, U.~Heintz, G.~Kukartsev, M.~Narain, {\em et~al.},
  ``{Heavy Higgs Scalars at Future Hadron Colliders (A Snowmass Whitepaper)},''
\href{http://arxiv.org/abs/1308.6334}{{\ttfamily arXiv:1308.6334 [hep-ex]}}.

\bibitem{Carena:2013qia}
M.~Carena, S.~Heinemeyer, O.~St\r{a}l, C.~Wagner, and G.~Weiglein, ``{MSSM
  Higgs Boson Searches at the LHC: Benchmark Scenarios after the Discovery of a
  Higgs-like Particle},''
  \href{http://dx.doi.org/10.1140/epjc/s10052-013-2552-1}{{\em Eur. Phys. J.}
  {\bfseries C73} (2013) 2552},
\href{http://arxiv.org/abs/1302.7033}{{\ttfamily arXiv:1302.7033 [hep-ph]}}.

\bibitem{Craig:2013hca}
N.~Craig, J.~Galloway, and S.~Thomas, ``{Searching for Signs of the Second
  Higgs Doublet},''
\href{http://arxiv.org/abs/1305.2424}{{\ttfamily arXiv:1305.2424 [hep-ph]}}.

\bibitem{Craig:2012pu}
N.~Craig, J.~A. Evans, R.~Gray, C.~Kilic, M.~Park, {\em et~al.},
  ``{Multi-Lepton Signals of Multiple Higgs Bosons},''
  \href{http://dx.doi.org/10.1007/JHEP02(2013)033}{{\em JHEP} {\bfseries 1302}
  (2013) 033},
\href{http://arxiv.org/abs/1210.0559}{{\ttfamily arXiv:1210.0559 [hep-ph]}}.

\bibitem{Eberhardt:2013uba}
O.~Eberhardt, U.~Nierste, and M.~Wiebusch, ``{Status of the two-Higgs-doublet
  model of type II},'' \href{http://dx.doi.org/10.1007/JHEP07(2013)118}{{\em
  JHEP} {\bfseries 1307} (2013) 118},
\href{http://arxiv.org/abs/1305.1649}{{\ttfamily arXiv:1305.1649 [hep-ph]}}.

\bibitem{Dumont:2014wha}
B.~Dumont, J.~F. Gunion, Y.~Jiang, and S.~Kraml, ``{Constraints on and future
  prospects for Two-Higgs-Doublet Models in light of the LHC Higgs signal},''
\href{http://arxiv.org/abs/1405.3584}{{\ttfamily arXiv:1405.3584 [hep-ph]}}.

\bibitem{Lee:2006wn}
S.~J. Lee, M.~Neubert, and G.~Paz, ``{Enhanced Non-local Power Corrections to
  the $\bar B\to X_s\gamma$ gamma Decay Rate},''
  \href{http://dx.doi.org/10.1103/PhysRevD.75.114005}{{\em Phys.Rev.}
  {\bfseries D75} (2007) 114005},
\href{http://arxiv.org/abs/hep-ph/0609224}{{\ttfamily arXiv:hep-ph/0609224
  [hep-ph]}}.

\bibitem{Benzke:2010js}
M.~Benzke, S.~J. Lee, M.~Neubert, and G.~Paz, ``{Factorization at Subleading
  Power and Irreducible Uncertainties in $\bar B\to X_s\gamma$ Decay},''
  \href{http://dx.doi.org/10.1007/JHEP08(2010)099}{{\em JHEP} {\bfseries 1008}
  (2010) 099},
\href{http://arxiv.org/abs/1003.5012}{{\ttfamily arXiv:1003.5012 [hep-ph]}}.

\bibitem{Bediaga:2012py}
{\bfseries LHCb} Collaboration, R.~Aaij {\em et~al.}, ``{Implications of LHCb
  measurements and future prospects},''
  \href{http://dx.doi.org/10.1140/epjc/s10052-013-2373-2}{{\em Eur.Phys.J.}
  {\bfseries C73} (2013) 2373},
\href{http://arxiv.org/abs/1208.3355}{{\ttfamily arXiv:1208.3355 [hep-ex]}}.

\bibitem{Dolan:2012ac}
M.~J. Dolan, C.~Englert, and M.~Spannowsky, ``{New Physics in LHC Higgs boson
  pair production},'' \href{http://dx.doi.org/10.1103/PhysRevD.87.055002}{{\em
  Phys.Rev.} {\bfseries D87} no.~5, (2013) 055002},
\href{http://arxiv.org/abs/1210.8166}{{\ttfamily arXiv:1210.8166 [hep-ph]}}.

\bibitem{Gupta:2013zza}
R.~S. Gupta, H.~Rzehak, and J.~D. Wells, ``{How well do we need to measure the
  Higgs boson mass and self-coupling?},''
  \href{http://dx.doi.org/10.1103/PhysRevD.88.055024}{{\em Phys.Rev.}
  {\bfseries D88} (2013) 055024},
\href{http://arxiv.org/abs/1305.6397}{{\ttfamily arXiv:1305.6397 [hep-ph]}}.

\bibitem{Ellwanger:2013ova}
U.~Ellwanger, ``{Higgs pair production in the NMSSM at the LHC},''
  \href{http://dx.doi.org/10.1007/JHEP08(2013)077}{{\em JHEP} {\bfseries 1308}
  (2013) 077},
\href{http://arxiv.org/abs/1306.5541}{{\ttfamily arXiv:1306.5541 [hep-ph]}}.

\bibitem{Barr:2013tda}
A.~J. Barr, M.~J. Dolan, C.~Englert, and M.~Spannowsky, ``{Di-Higgs final
  states augMT2ed -- selecting $hh$ events at the high luminosity LHC},''
\href{http://arxiv.org/abs/1309.6318}{{\ttfamily arXiv:1309.6318 [hep-ph]}}.

\bibitem{Dolan:2013rja}
M.~J. Dolan, C.~Englert, N.~Greiner, and M.~Spannowsky, ``{Further on up the
  road: $hhjj$ production at the LHC},''
  \href{http://dx.doi.org/10.1103/PhysRevLett.112.101802}{{\em Phys.Rev.Lett.}
  {\bfseries 112} (2014) 101802},
\href{http://arxiv.org/abs/1310.1084}{{\ttfamily arXiv:1310.1084 [hep-ph]}}.

\bibitem{No:2013wsa}
J.~M. No and M.~Ramsey-Musolf, ``{Probing the Higgs Portal at the LHC Through
  Resonant di-Higgs Production},''
\href{http://arxiv.org/abs/1310.6035}{{\ttfamily arXiv:1310.6035 [hep-ph]}}.

\bibitem{Baglio:2014nea}
J.~Baglio, O.~Eberhardt, U.~Nierste, and M.~Wiebusch, ``{Benchmarks for Higgs
  Pair Production and Heavy Higgs Searches in the Two-Higgs-Doublet Model of
  Type II},''
\href{http://arxiv.org/abs/1403.1264}{{\ttfamily arXiv:1403.1264 [hep-ph]}}.

\bibitem{Chen:2014xra}
C.-R. Chen and I.~Low, ``{A Double Take on New Physics in Double Higgs
  Production},''
\href{http://arxiv.org/abs/1405.7040}{{\ttfamily arXiv:1405.7040 [hep-ph]}}.

\end{thebibliography}\endgroup
\bibliographystyle{utphys}
}
\end{document}